\documentclass[sigconf]{acmart}
\AtBeginDocument{%
  \providecommand\BibTeX{{%
    \normalfont B\kern-0.5em{\scshape i\kern-0.25em b}\kern-0.8em\TeX}}}

\copyrightyear{2021}
\acmYear{2021}
\setcopyright{rightsretained}
\acmConference[CHI '21]{CHI Conference on Human Factors in Computing Systems}{May 8--13, 2021}{Yokohama, Japan}
\acmBooktitle{CHI Conference on Human Factors in Computing Systems (CHI '21), May 8--13, 2021, Yokohama, Japan}\acmDOI{10.1145/3411764.3445186} \acmISBN{978-1-4503-8096-6/21/05}

\usepackage{xspace,xpunctuate}
\usepackage{changepage}
\newcommand{\ie}{{i.e.,}\xspace}
\newcommand{\eg}{{e.g.,}\xspace}
\newcommand{\etal}{{et~al\xperiod}\xspace}

\newcommand{\placeholder}{\textcolor{black}} %
\newcommand{\camerareadychange}{\textcolor{black}}

\usepackage{changepage}
\newcommand{\quotateblock}[2]{%
\vspace{0.3em}
\begin{adjustwidth}{1.0em}{0.5em}%
#1:\textit{``#2'’}%
\end{adjustwidth}%
\vspace{0.3em}
}

\begin{document}

\title{\camerareadychange{Screen Recognition:} Creating Accessibility Metadata for Mobile Applications from Pixels}

\author{Xiaoyi Zhang}
\affiliation{Apple Inc.}
\email{xiaoyiz@apple.com}

\author{Lilian de Greef}
\affiliation{Apple Inc.}
\email{ldegreef@apple.com}

\author{Amanda Swearngin}
\affiliation{Apple Inc.}
\email{aswearngin@apple.com}

\author{Samuel White}
\affiliation{Apple Inc.}
\email{samuel_white@apple.com}

\author{Kyle Murray}
\affiliation{Apple Inc.}
\email{kyle_murray@apple.com}

\author{Lisa Yu}
\affiliation{Apple Inc.}
\email{lixiu_yu@apple.com}

\author{Qi Shan}
\affiliation{Apple Inc.}
\email{qshan@apple.com}

\author{Jeffrey Nichols}
\affiliation{Apple Inc.}
\email{jwnichols@apple.com}

\author{Jason Wu}
\affiliation{Apple Inc.}
\email{jason_wu2@apple.com}

\author{Chris Fleizach}
\affiliation{Apple Inc.}
\email{cfleizach@apple.com}

\author{Aaron Everitt}
\affiliation{Apple Inc.}
\email{aeveritt@apple.com}

\author{Jeffrey P. Bigham}
\affiliation{Apple Inc.}
\email{jbigham@apple.com}

\renewcommand{\shortauthors}{Zhang, de Greef, Swearngin, et al.}

\begin{abstract}
Many accessibility features available on mobile platforms require applications (apps) to provide complete and accurate metadata describing user interface (UI) components. Unfortunately, many apps do not provide sufficient metadata for accessibility features to work as expected. In this paper, we explore inferring accessibility metadata for mobile apps from their pixels, as the visual interfaces often best reflect an app's full functionality. We trained a robust, fast, memory-efficient, on-device model to detect UI elements using a dataset of \placeholder{77,637} screens \camerareadychange{(from 4,068 iPhone apps)} that we collected and annotated. To further improve UI detections and add semantic information, we introduced heuristics (\eg UI grouping and ordering) and additional models (\eg recognize UI content, state, interactivity). We built \camerareadychange{Screen Recognition} to generate accessibility metadata to augment iOS VoiceOver. In a study with \placeholder{9} screen reader users, we validated that our approach improves the accessibility of existing mobile apps, enabling even previously inaccessible apps to be used.

\end{abstract}

\begin{CCSXML}
<ccs2012>
   <concept>
       <concept_id>10003120.10011738.10011775</concept_id>
       <concept_desc>Human-centered computing~Accessibility technologies</concept_desc>
       <concept_significance>500</concept_significance>
       </concept>
 </ccs2012>
\end{CCSXML}

\ccsdesc[500]{Human-centered computing~Accessibility technologies}

\keywords{mobile accessibility, accessibility enhancement, ui detection}

\maketitle

\section{Introduction}

\begin{figure*}[t]
  \centering
  \includegraphics[width=\textwidth]{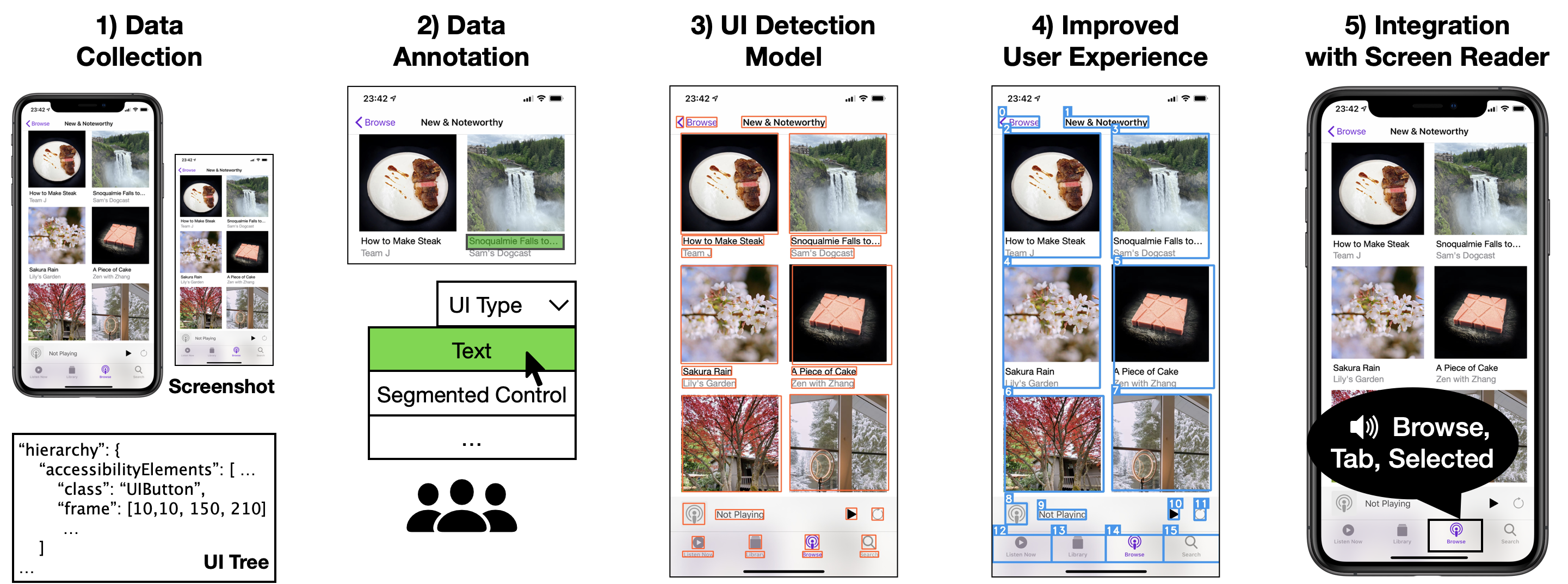}
  \caption{Overview of our approach: 1) Ten workers collected screenshots and UI trees in apps. 2) Forty workers annotated the screens to provide complete metadata. 3) An on-device model (trained with annotated data) detects UI elements from pixels. 4) Heuristics (\eg ordering and grouping) and additional models (\eg OCR) augment UI detections to improve the user experience. 5) The generated accessibility metadata is made available to screen readers for use at runtime to improve the experienced accessibility of apps.}
  \Description{Our system overview diagram shows each distinct phase of our work. The first is 1) data collection which shows a mobile phone with a screenshot and UI tree being captured from it. 2) Data annotation shows part of this screenshot with one UI element annotated and a dropdown UI type selection list with Container selected, and an icon to represent crowd workers. 3) UI detection model shows the screenshot with red boxes around each UI element detected by the model. 4) Improved user experience shows the same screenshot with blue boxes around each UI element, some of which from the previous screenshot have been grouped together. The screenshot also shows the annotated navigation order. 5) Integration with iOS VoiceOver shows a mobile phone screen with a UI element highlighted by VoiceOver with a speech bubble showing the text "Browse, Tab, Selected" to represent the VoiceOver speaking the UI element description. } 
  \label{fig:system_diagram}
\end{figure*}

Many accessibility features (\eg screen readers \cite{appleVoiceOver, googleTalkback}, switch control \cite{switchcontrol,switchaccess}) only work on apps that provide a complete and accurate description of their UI semantics (\eg class="TabButton", state="Selected", alternative text="Profile"). Despite decades of work on developer tools and education to support accessibility across many different platforms \cite{pearson2011tool, checker2011idi, carter2007techniques}, apps still do not universally supply accessibility metadata on any platform \cite{hanson2013progress}. For example, all of the 100 most-downloaded Android apps had basic accessibility issues in a recent study \cite{ross2017epidemiology}.
The lack of appropriate metadata needed for accessibility features is a long-standing challenge in accessibility.

Apps may not be fully accessible for a variety of reasons. Developers may be unaware of accessibility, lack necessary accessibility expertise to make their apps accessible, or deprioritize accessibility. Developers often use third-party UI toolkits that can work across platforms, but have limited built-in accessibility support.
For example, to make Unity apps accessible non-visually, developers either have to use a paid Unity plugin \cite{unityPlugin} that replicates a screen reader experience (catches gestures on a full-screen overlay and announces focused UI elements) or manually recreate a similar accessible experience from scratch. First party UI toolkits (\eg UIKit for iOS) make accessibility much easier, but app developers still need to provide accessibility metadata, such as alternative text or accessibility elements for custom views \cite{uiaccessibility}. Much of the focus is on creating new content and apps that are accessible, although most platforms have a substantial legacy of inaccessible content and apps that may no longer have active developer support. The myriad challenges of achieving accessibility across many different platforms and toolkits for so many different apps motivated us to automatically create accessibility metadata from the pixels of app user interfaces.

In this paper, we introduce a new method for providing accessibility metadata automatically from the pixels of the visual user interface. In practice, the visual interfaces for mobile apps often receive the most attention from developers and best represent an app's intended functionality.
To facilitate this method, we collected, annotated, and analyzed \placeholder{77,637} screens \camerareadychange{(from 4,068 iPhone apps)}. Using this dataset, we trained a robust, fast, memory-efficient, on-device object detection model to extract UI elements from screenshots, which achieves \placeholder{71.3\%} Mean Average Precision. To add semantics and further improve UI detection results, we introduced heuristics to correct detections, provide navigation order, and group relevant detections; we also used additional models to recognize UI content, selection states, and interactivity. Our method creates metadata useful for a wide variety of accessibility services \cite{appleVoiceOver, googleTalkback, switchcontrol, switchaccess, voicecontrol}, and can be used either alone or to improve existing accessibility metadata in an app. Fig. \ref{fig:system_diagram} shows integration with screen readers as an example.

To demonstrate the utility of our approach, we created \camerareadychange{Screen Recognition} that provides this metadata to iOS VoiceOver, using only screenshot pixels as input. In a user study with \placeholder{9} screen reader users, we validated that participants could use a wide variety of existing mobile apps, including previously inaccessible ones. We also explored the limitations of our approach, and discussed future improvements as well as potential integration with developer tools. At the high level, we believe our approach illustrates a new way to start addressing the long-standing challenge of inaccessible apps on mobile and, eventually, other platforms.

Our work contributes:
\begin{itemize}
\item An analysis of the characteristics (\eg UI distribution, accessibility issues) of a large dataset of \placeholder{77,637} screens \camerareadychange{(from 4,068 iPhone apps)} we collected and annotated.
\item A robust, fast, memory-efficient, on-device object detection model to extract UI elements from raw pixels in a screenshot, which we trained and evaluated.
\item Augmentation of UI detections for a better user experience. Our heuristics correct detections, provide navigation order, and group relevant detections; additional models recognize UI content, state, and interactivity.
\item A user study with \placeholder{9} screen reader users, who tried out apps of their choices with VoiceOver with and without our approach applied. Their feedback validates our approach can significantly improve accessibility on existing apps.
\end{itemize}

\section{Related Work}

Our work builds from prior work on {\em (i)} supporting mobile accessibility, {\em (ii)} detecting UI elements from pixels, and {\em (iii)} automatic understanding of UI semantics.

\subsection{Supporting Mobile Accessibility}

Most mobile platforms (\eg iOS, Android) these days contain built-in accessibility services like screen readers \cite{appleVoiceOver, googleTalkback}) and support for accessible input \cite{switchaccess, switchcontrol}, which allows people to use a wide variety of abilities to use them. These technologies depend heavily on the availability of accessibility metadata provided by developers to expose the underlying semantics of apps \cite{iosax, androidax}. When this metadata is provided, accessibility services have programmatic access to what UI elements are available, what their content contains, what state they are in, and what interactions can be performed on them. As such, this metadata is fundamental to enabling accessibility services to change the modalities by which someone can interact with UIs and support people using them in different ways. However, both previous research \cite{ross2017epidemiology, ross2018epidemiology, ross2020epidemiology}, and our analysis in this paper (Section \ref{sec:datasetAnalysis}), show that developers routinely fail to include this information, and many mobile apps remain inaccessible.

Prior approaches for addressing this problem include encouraging developers to fix accessibility problems through education \cite{carter2007techniques}, standards \cite{power2012guidelines,wcag}, and better testing tools \cite{checker2011idi,brajnik2008comparative}. While these approaches are important, since ultimately developers will be able to best make their apps accessible, current progress remains slow \cite{hanson2013progress}, motivating research that tries to solve the problem without relying on the original developers' cooperation. Supplanting or automatically generating metadata for existing inaccessible apps may provide more immediate benefits for users. To this end, Interaction Proxies allows missing metadata to be supplanted by end-users by manually labeling UI elements and forming a shared repository of such information for runtime repair \cite{zhang2017interaction, zhang2018robust}. However, this approach requires active volunteers to update and maintain annotations and customizations for a large number of apps which are frequently updated. The largest prior attempt to ``crowdsource'' accessibility, Social Accessibility for the Web \camerareadychange{\cite{takagi2008social}}, showed early promise but ultimately was unable to make a big dent in the problem of Web accessibility.

Previous research in automated image description generation \cite{guinness2018caption, gleason2020twitter} and UI label prediction \cite{chen2020unblind} also creates specific types of accessibility automatically metadata from pixels. However, it only comprises a subset of the information needed for accessibility services to work as expected (\ie these approaches do not make missing UI elements available). Our approach goes much further toward complete accessibility metadata generation from pixels, automatically inferring missing UI elements (size, position, type, content, and state), proper navigation order, and grouping.

\subsection{UI Detection from Pixels}
Our work involves inferring the locations and semantics of UI elements on a screen to provide metadata for accessibility services. Some early work extracted UI information to modify interfaces from non-pixel sources, such as the accessibility~\cite{changeAssociating2011} and instrumentation APIs~\cite{olsen_implementing_1999, eagan_cracking_2011}, or software tutorial videos~\cite{banovicWaken2012}. However, they require API access to enable instrumentation or to expose UI elements, and often have incomplete access to UI metadata. Pixel-based approaches, in contrast, are more independent of the underlying app toolkit and do not require UIs to be exposed via APIs.

Using pixels to identify UI elements has long been used to make apps accessible. For instance, the OutSpoken screen reader for Windows 3.1 allowed users to label icons on the screen, which it then recognizes from their pixels alone \cite{outspoken}. Inferring information from pixels of interfaces has been applied in diverse applications such as interface augmentation and remapping \cite{dixon2010prefab, zettlemoyer1999visual, banovicWaken2012, changeAssociating2011}, GUI testing~\cite{yeh2009sikuli}, data-driven design for GUI search~\cite{chen2019gallery, chen2020wireframesearch, huang2019swire} or prototyping~\cite{swearngin2018rewire}, generating UI code from existing apps to support app development~\cite{nguyen2015remaui, beltramelli2017pix2code, chen2018neural, moran2018machine, chen2020object}, and GUI security~\cite{chen2019gui}. Some work also employed pixel-based methods to improve accessibility, such as Prefab, which augments existing app interface with target-aware pointing techniques that enhance interaction for people with motor impairments ~\cite{dixon2012bubblecursor}.

There are multiple approaches to pixel-based interpretation of interfaces. Recent work by Chen \etal~\cite{chen2020object} categorizes and evaluates two major GUI detection approaches: using traditional image processing methods (\eg edge/contour detection \cite{nguyen2015remaui}, template matching ~\cite{dixon2010prefab, yeh2009sikuli, pongnumkul2011pause}) and using deep learning models trained on large-scale GUI data ~\cite{chen2018neural, chen2020object}.

\textbf{Traditional image processing methods:} Edge/contour detection methods \cite{nguyen2015remaui, swearngin2018rewire, tech2019ui2code} detect and merge edges into UI elements, which can work well on simple GUIs, but can fall short on images (\eg gradient background, photos) or complex GUI layouts. Alternatively, template matching methods ~\cite{yeh2009sikuli, dixon2010prefab, zettlemoyer1999visual} may work better on these cases as there are more features in complex UI elements. However, they require feature engineering and templates to recognize the shape and type of UI elements, which may limit them to detecting UIs without much variance in visual features.

\textbf{Deep learning models:} Pix2Code \cite{beltramelli2017pix2code} applies an end-to-end neural image captioning model to predict a Domain Specific Language description of the interface layout. Chen \etal \cite{chen2018neural} applies a similar CNN-RNN model to generate a UI skeleton (consisting of widget types and layout) from a screenshot. In both works, the generated layouts only provide relative locations of UI elements in the view hierachy, while most accessibility services require the absolute location of each UI element on the screen. Object detection models can also locate UI elements on a screen: GalleryDC \cite{chen2019gallery} detects UIs with Faster RCNN model ~\cite{ren2015faster} to auto-create a gallery of GUI design components, and White \etal~\cite{white2019improving} applies YOLOv2 model \cite{redmon2016you} to identify GUI widgets in screenshots to improve GUI testing.

\textbf{Hybrid methods:} Some work~\cite{moran2018machine, chen2019automated, chen2020object} first uses traditional image processing methods (\eg edge detection) to locate UI elements, and then applies CNN-based classification to determine the semantics of UI elements (\eg UI type).

Many of these approaches aim at facilitating the early stages of app design rather than accessibility; therefore, they only focus on a subset of metadata (\ie UI location and type) needed by accessibility services. In contrast, our deep learning-based approach is optimized for accessibility use cases, provides more comprehensive metadata from pixels (\eg UI state, navigation order, grouping), and requires less resources to efficiently run on mobile phones.

\subsection{Understanding UI Semantics}
Beyond detecting UI elements from a screenshot, our work requires obtaining a more thorough understanding of UI semantics (\eg UI type, state, navigation order, grouping, image and icon descriptions), which can further improve accessibility services and user experience. Beyond UI element detection, recent data-driven design work relies on large UI design datasets to infer higher level interface semantics, such as design patterns~\cite{nguyen2018patterns}, flows~\cite{deka2016erica}, and tags~\cite{chen2020tags}. Screen readers require inferring a different set of semantics, such as the navigation order and UI hierarchy (\ie grouping) for a screen. Some work applies machine learning on UI design datasets to infer interface layout~\cite{kumar2011bricolage, bielik2018robust, chen2018neural, chen2020wireframesearch}. Other work extracts interface structure through system APIs~\cite{memon2003gui}. In contrast, we worked with people with visual impairments to create a heuristics-based approach to group and provide a navigation order for detected UI elements to better fit the expected user experience of a screen reader. 

In addition, it is important for accessibility services to know the state and interactivity of an element (\eg clickability, selection state), so they can handle elements differently (\eg VoiceOver will announce "Button" for clickable UI elements; Switch Control may skip non-interactive UI elements in navigation to save time). The work by Swearngin \etal~\cite{swearngin2018tappability} predicts the human perception of the tappability of mobile app UI elements, based on human annotations and common clickability signifiers (\eg UI type, size). In our work, we built and integrated a model to predict the clickability of some UI elements based on similar features (\ie, size, location, icon type); however, our goal is to predict the actual clickability of UI elements rather than the human perception; as~\cite{swearngin2018tappability} shows, these two often mismatch. Beyond previous work to detect UI elements from screenshots~\cite{moran2018machine, beltramelli2017pix2code, chen2020object, nguyen2015remaui}, we additionally detect the selection state for relevant UI element types (\eg Toggle, Checkbox) by inferring additional UI element type classes. 

Lastly, we infer semantic information from UI elements to improve the screen reader experience. To recognize icons, previous work ~\cite{liu2018semantics, xi2019deepintent, xiao2019icon} trained image classification models from UI design datasets~\cite{deka2017rico}. To describe content in pictures, prior work used deep learning models to generate natural language descriptions of images ~\cite{karpathy2015deep,hossain2019comprehensive}, and some accessibility improvement research has also leveraged crowdsourcing to generate image captions ~\cite{guinness2018caption, gleason2020twitter, gurari2020captioning}. We use an existing Icon Recognition engine and Image Descriptions feature in iOS \cite{ios14preview} to generate alternative text for detected icons and pictures, respectively.

\section{\texorpdfstring{\MakeLowercase{i}}{i}OS App Screen Dataset}

To train and test our UI detection model, we first created a dataset of \placeholder{77,637} screens from \placeholder{4,068} iPhone apps. Creating our dataset required two steps: {\em (i)} collecting the screens, in which we captured screenshots and extracted information of their UI trees and accessibility trees, and {\em (ii)} manually annotating the visual UI elements on the screens. We fine-tuned this process, including our custom data collection and annotation software, through several pilot trials. \camerareadychange{The detail of workers we recruited in data collection and annotation and the instructions for annotators are in supplementary material.}

\subsection{Screen Collection}

We collected a total of \placeholder{80,945} unique screens from \placeholder{4,239} iPhone apps, manually downloading and installing the top 200 most-downloaded free apps for each of 23 app categories (excluding Games and AR Apps) presented in the U.S. App Store. Note that some apps are in multiple categories. We skipped apps that required credit card information (\eg for in-app payment) or sensitive personal information (\eg log-in for banking apps, personal health data). We created a set of placeholder accounts to log into apps that required this in order to access more of their screens. \camerareadychange{We collected all screens between 12/09/2019 and 02/13/2020.}

Ten workers manually traversed screens in each app to collect screen information using custom software we built. For each traversed screen, our software collected a screenshot and metadata about UI elements. The metadata includes a tree structure of all exposed UI elements on a screen, and properties of each UI element (\eg class, bounding box, trait). We also extracted accessibility properties of each UI element that are exposed to the Accessibility APIs (\eg isAccessibilityElement, hint, value). Our collection software also assigned a unique identifier to each UI element. The collected metadata has the same limitations in terms of completeness and correctness that motivated our approach. Therefore, establishing a reliable ground truth requires annotating the collected screens.

\begin{figure}[t]
  \centering
  \includegraphics[width=\linewidth]{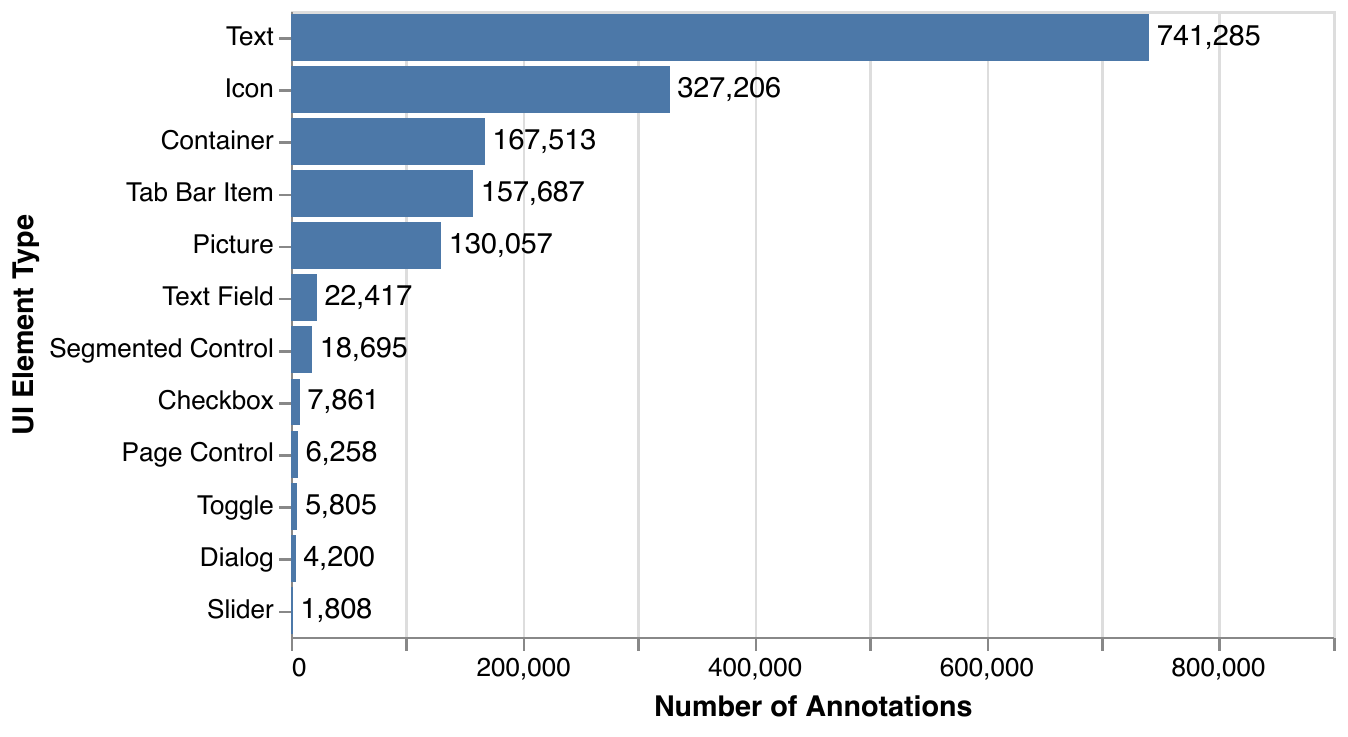}
  \caption{The number of annotations for each UI element type represented in the dataset.}
  \Description{A bar chart displaying the number of annotations for each UI element type in the dataset. "Text" has by far the highest representation, with 741285 annotations. The other elements, ordered by their number of annotations, are: Text (741285), Icon (327206), Container (167513), Tab Bar Item (157687), Picture (130057), Text Field (22417), Segmented Control (18695), Checkbox (7861), Page Control (6258), Toggle (5805), Dialog (4200), and Slider (1808).
  }
  \label{fig:uielement_representation}
\end{figure}
\subsection{Screen Annotation} \label{sec:annotations}

Forty workers annotated all visually discernible UI elements in the screenshots.  The annotation process had two components: {\em (i)} segmentation and {\em (ii)} classification.

In segmentation, workers determined a bounding box for each UI element. The annotation tool helped to direct workers toward using a bounding box provided by a captured UI element, when an appropriate one was available, to improve consistency across workers and annotations. When no bounding box was available for a UI element, the annotators manually drew a box. We established guidelines to ensure consistency and quality, such as how to draw a tight bounding box; when to group or separate bounding boxes; what elements to ignore (\eg background); and how to handle edge cases (\eg clipped elements).

For classification, workers assigned attributes to the identified UI elements. For each, annotators assigned one of 12 common UI types based on visual inspection: Checkbox, Container, Dialog, Icon, Picture, Page Control, Segmented Control, Slider, Text, Text Field, Tab Bar Item, and Toggle. Visual examples of UI types are available in our Supplementary Material. Two authors of this paper (a researcher and a senior accessibility engineer) chose these UI types by examining \placeholder{500} samples screens to identify which elements are important for accessibility services. For UI elements not in one of these UI types, such as advertisements and map views, annotators marked them as Other (0.5\% of all annotations). Annotators also labeled additional information for specific UI types, such as whether each Checkbox or Toggle is selected, whether an element is clickable, and which part of an element the bounding box encapsulates (\eg a Text Field's outline or its placeholder text). As the attributes can sometimes be ambiguous, annotators could also choose the option ``unsure''.

To monitor data quality and iteratively correct operational errors, two researchers examined a random selection of 6\% of the screens after each of the 20 batches was annotated.
In addition, our Quality Assurance (QA) team examined 20\% of screens that were annotated in each batch and shared a detailed QA report. Based on this, we worked with the annotators to reduce common errors. The total error rate we observed was \placeholder{2.62\%} in all batches, which dropped from \placeholder{4.49\%} (first batch) to \placeholder{1.27\%} (last batch). On average, the error rate of bounding boxes (\eg missing or extra annotation, too big or small of a box) was \placeholder{1.35\%}; the error rate of UI type was \placeholder{0.54\%}; and, the error rate of UI attributes (\eg clickable/not clickable, selected/not selected) was \placeholder{0.73\%}. We found that the primary errors were missing annotations for pictures and confusion between clickable and not-clickable elements, a known challenge \cite{swearngin2018tappability}.

During the annotation phase, we discarded \placeholder{3,308} screens that were accidentally captured when an app was in the transition between screens, accidentally showed an AssistiveTouch button, would take too long to annotate (\eg contain 50+ UI Elements), or had much non-English text. All of these screens represent opportunities for future work, but we considered them out of scope for the first version of the project. This resulted in \placeholder{77,637} total annotated screens.

\subsection{Dataset Composition Analysis} \label{sec:datasetAnalysis}

To better understand the composition of our dataset, we conducted two analyses. The first analysis explores our dataset's biases between different UI types, which may impact our model performance. The annotations revealed an imbalance of UI types in app screens, as shown in Fig. \ref{fig:uielement_representation}: Text has the highest representation (\placeholder{741,285} annotations); Sliders have the lowest (\placeholder{1,808} annotations); and the top 4 UI types comprise 95\% of annotations. We consider such data imbalances in model training to improve performance on underrepresented UI types, as discussed in Section \ref{sec:modelArchitecture}.

Our second analysis examined discrepancies between annotations and UI element metadata for each screen to estimate how many UI elements were not available to accessibility services. Note that there are important limitations to this estimate, as these annotations are not a perfect reflection of all UI elements that are important to accessibility service: annotators sometimes ignore UI elements; annotations may include UI elements created for aesthetic reasons that are not important for accessibility; annotations may have additional errors as mentioned in Section \ref{sec:annotations}. Furthermore, the available UI element metadata do not account for all possible ways to make an app accessible. For instance, some apps use custom methods (\eg re-implement a screen reader) when their UI toolkits do not support native accessibility. Our analysis focused on UI elements exposed to the native screen reader with the understanding that annotations from human observation are still a close approximation to important accessibility elements. In most cases, the differences between annotations and existing accessibility metadata suggest potential accessibility issues in a screen.

Matching annotations to UI elements in the view hierarchy presents several challenges. Annotation bounding boxes sometimes do not line up exactly to the underlying UI element's. Sometimes they differ in the way they partition an overarching UI component (\eg treating individual lines or paragraphs of Text as individual elements versus one large piece of Text, or exposing separate elements for the Icon, Container, and Text of a button versus just exposing one of these parts to accessibility services). Some of the larger exposed accessibility elements also have many smaller accessibility elements nested inside of them, creating further ambiguity for which elements match to which annotations.

To make this comparison, we ignored advertisements or maps and examined the remaining annotations with the following procedure (which we describe in more detail in this paper's supplemental material): categorize annotations by whether any bounding box \verb|B| from a non-fullscreen accessibility element contains or overlaps with annotation's bounding box \verb|A|, where we define "containment" as whether at least \placeholder{85\%} of \verb|A|'s area lies within \verb|B|, and "overlap" as whether the intersection over union of \verb|A| and \verb|B| is at least \placeholder{5\%}. When an exposed element's bounding box contains only one annotation, we consider them as a match with each other. %
We consider annotations that have no overlap with any exposed accessibility elements to not have a match, with a few exceptions like Icons that horizontally align with already-matched Text elements that are not independently clickable, as is often the case in table cells or buttons. %
For the remaining possibilities — an exposed element contains more than one annotation, or overlaps but does not contain an annotation — we examined \placeholder{150} example screens with different UI element types to develop heuristics, which we tested on another \placeholder{150} example screens. 
These heuristics try to match inconsistent bounding boxes for Pictures and Text elements, nested bounding boxes, and functionally redundant annotations.
To maintain a high recall on matches, these heuristics have lenient tolerances and do not require clickable annotations to match with clickable UI elements. Hence, we may have underestimated the number of un-matched annotations.

\begin{figure}[t]
  \centering
  \includegraphics[width=\linewidth]{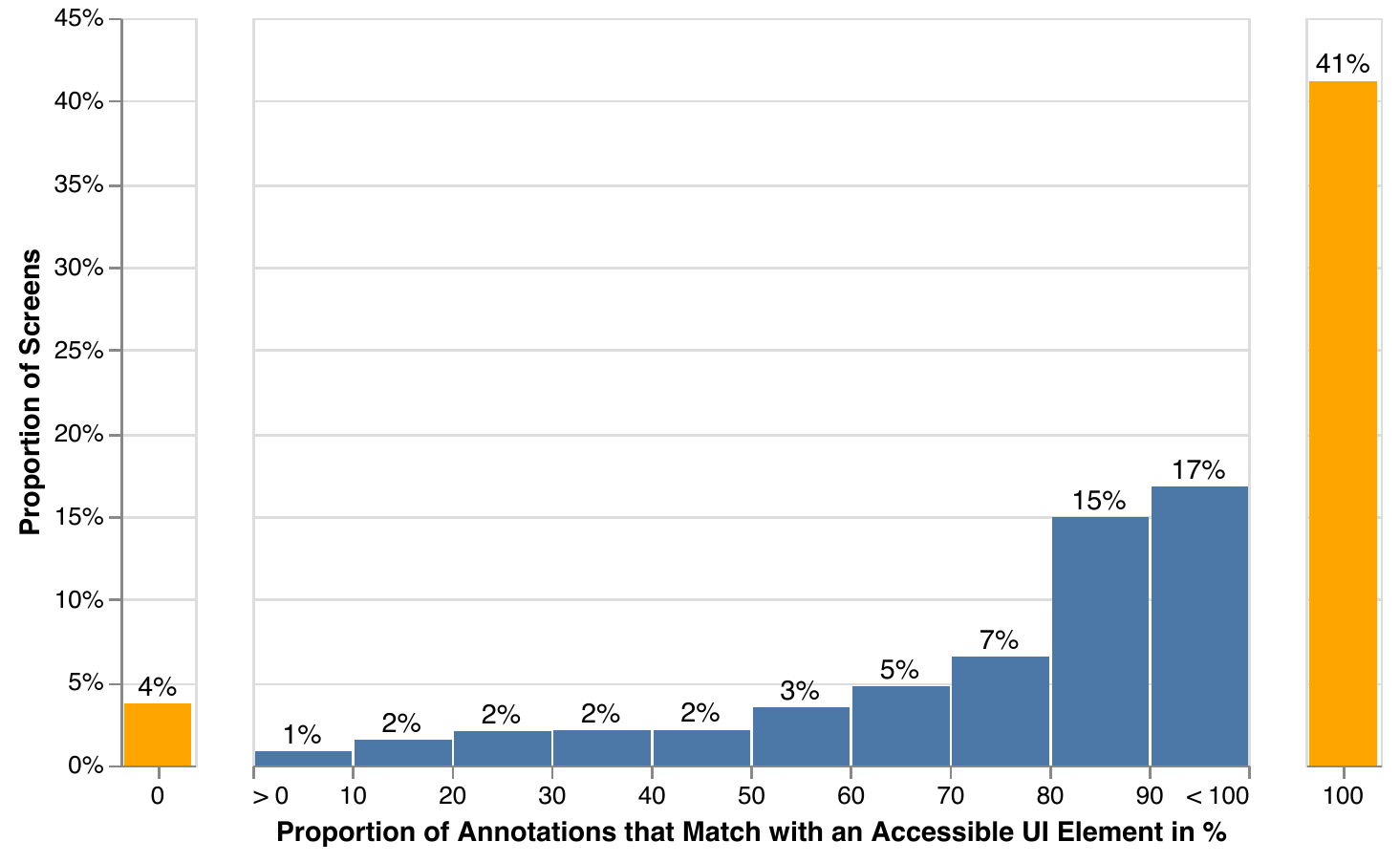}
  \caption{A histogram of the percentage of screens binned by what percentage of their annotations matched with an existing accessible UI element, with separate bars to show the percentage of screens with 0\% matches and 100\% matches.}
  \Description{For \placeholder{4}\% of screens, none of the annotations match with accessibility elements.
      \placeholder{41}\% of screens have matches for all annotations.
      For the remaining \placeholder{55}\% of screens that have matches for some but not all of their annotations, the number of screens increases with the percentage of matches. A full breakdown is as follows: 0\% matches between annotations and screen accessibility elements: 4\% of screens. more than 0 to 10\% matches: 1\% of screens. 10 to 20\% matches: 2\% of screens. 30 to 40\% matches: 2\% of screens. 40 to 50\% matches: 2\% of screens. 50 to 60\% matches: 3\% of screens. 60 to 70\% matches: 5\% of screens. 70 to 80\% matches: 7\% of screens. 80 to 90\% matches: 18\% of screens. 90 to less than 100\% matches: 17\% of screens. 100\% matches: 41\% of screens.}
  \label{fig:missing-element-histogram}
\end{figure}

\begin{figure}[t]
  \centering
  \includegraphics[width=\linewidth]{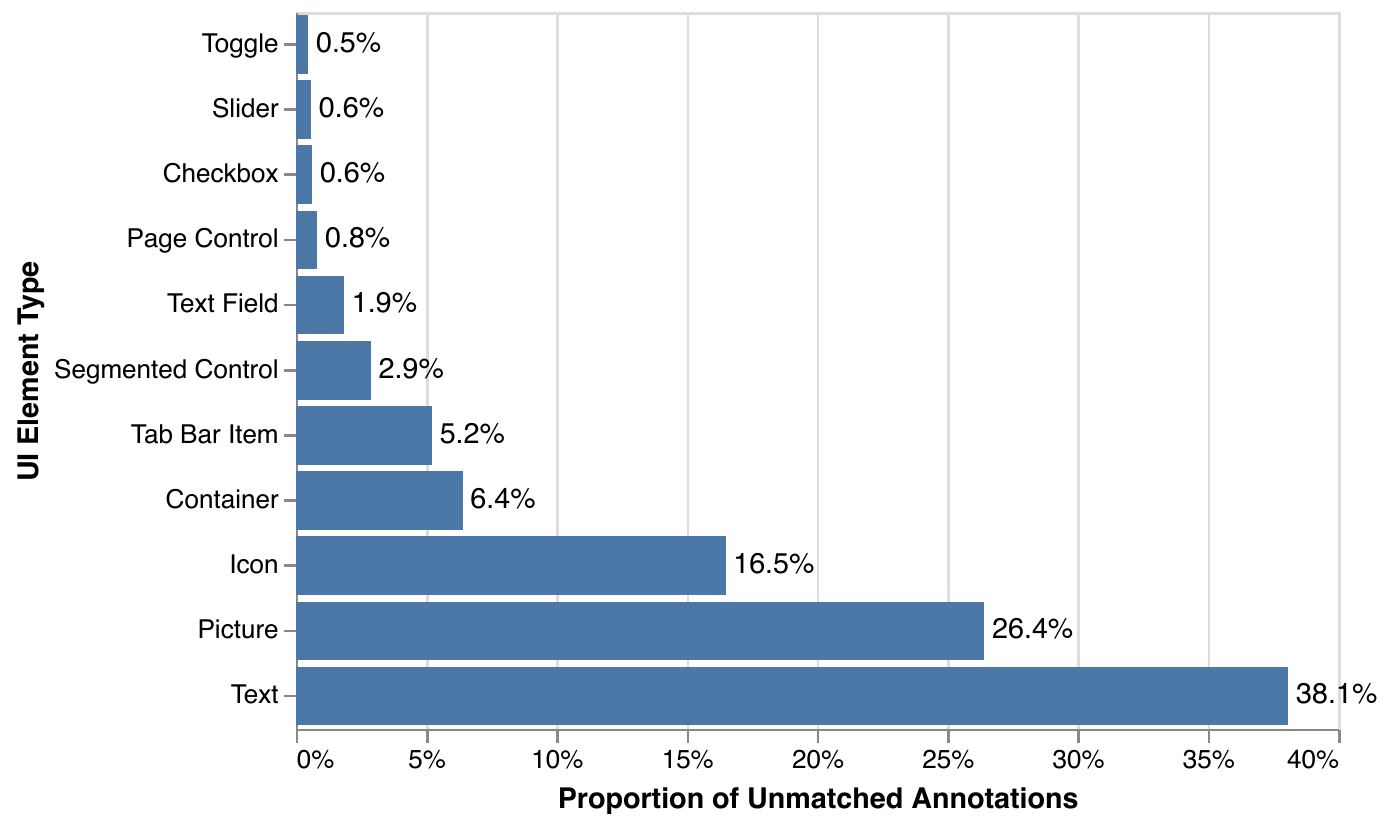}
  \caption{A bar chart showing the breakdown of unmatched annotations by UI Element Type.}
  \Description{A bar chart showing the representation of different UI element types among unmatched annotations. Text annotations have the highest representation, making up \placeholder{38.1\%} of unmatched annotations, followed by 
      Picture (\placeholder{26.4\%}, 
      Icon (\placeholder{16.5\%}), 
      Container (\placeholder{6.4\%}), 
      Tab Bar Item (\placeholder{5.2\%}), 
      Segmented Control (\placeholder{2.9\%}), 
      Text Field (\placeholder{1.9\%}), 
      Page Control (\placeholder{0.8\%}), 
      Checkbox (\placeholder{0.6\%}), 
      Slider (\placeholder{0.6\%}), 
      and Toggle (\placeholder{0.5\%}).}
  \label{fig:missing-annotation-percentages}
\end{figure}

As shown in Fig. \camerareadychange{\ref{fig:missing-element-histogram}}, \placeholder{59\%} of screens have annotations that do not match to any accessible UI element (a mean of \placeholder{5.3} and median of \placeholder{2} annotations per screen). \camerareadychange{We found that 94\% of the apps in our dataset have at least one such screen, rather than just a few unusable apps contributing to a large number of these screens.} In these screens, our accessibility metadata generation approach has the potential to expose missing accessibility elements. Our approach could be especially useful in providing accessibility metadata for the \placeholder{4\%} of screens that do not expose any accessibility elements. As shown in Fig. \camerareadychange{\ref{fig:missing-annotation-percentages}}, the prevalence of matches between annotations and accessible UI elements varied by UI type. Of all unmatched annotations, \placeholder{33\% are Text and 21\% are Picture}. The discrepancy between annotations and existing accessibility metadata demonstrates the importance of annotation, which provides additional information compared to prior mobile UI collection datasets \cite{deka2017rico}.

\section{UI Detection Model}
We trained an object detection model to extract UI elements from pixels, which locates UI elements on an app screenshot and predicts their UI types. After experimenting with a variety of network architectures and configurations, we chose one that balances accuracy and on-device performance. We describe the model architecture, data, and evaluation below.

\subsection{Model Architecture} \label{sec:modelArchitecture}
We started by experimenting with Faster R-CNN (and its extension Mask R-CNN)~\cite{ren2015faster, he2017mask} which is one of the best-performance object detection models evaluated on public datasets. As the goal is to run this detection model on-device (iPhone), the inference time and memory footprint became limiting factors. Faster R-CNN takes more than 1 second for each screen and more than 120 MB memory, unsuitable for on-device inference. We then experimented with TuriCreate Object Detection toolkits~\cite{turicreate}, which is easy to use and optimized for iOS devices. The TuriCreate model uses approximately half of the memory (60M) and has a significantly faster inference time (around 20ms). In the pursuit of a more efficient model with tighter bounding box predictions and a higher mAP (mean Average Precision), we converged on a SSD (Single Shot MultiBox Detector) model~\cite{liu2016ssd} that meets our requirements. Specifically, we used MobileNetV1 (instead of large ResNet) as a backbone to reduce memory usage. Another common challenge in object detection tasks is detecting small objects. Unfortunately, UI elements are relatively small compared to most targets often seen in object detection tasks. We used Feature Pyramid Network (FPN), which is a feature extractor designed with a hierarchical pyramid to improve accuracy and speed when detecting objects in different scales. To handle class-imbalanced data, discussed in Section \ref{sec:datasetAnalysis}, we performed data augmentation on underrepresented UI types during training, and applied a class-balanced loss function (more weight for underrepresented UI types). Our final architecture uses only 20MB of memory (CoreML model), and takes only about 10ms for inference \camerareadychange{(on an iPhone 11 running iOS 14)}. To train this model, we used \placeholder{4} Tesla V100 GPUs for \placeholder{20} hours (\placeholder{557,000} iterations).

\begin{figure}[t]
  \centering
  \includegraphics[width=\linewidth]{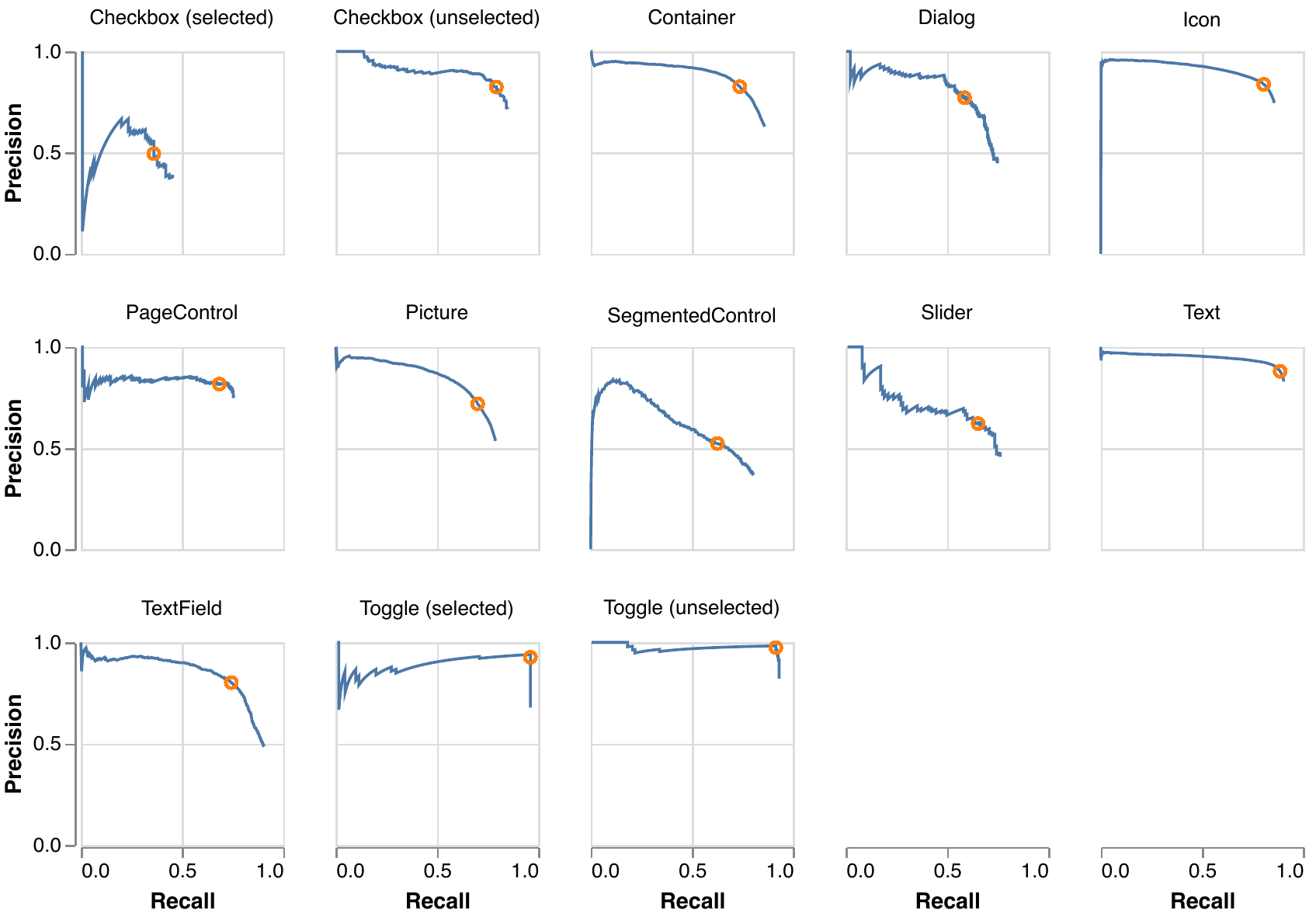}
  \caption{Our model's precision-recall curves for detecting each UI type. Examining individual performance enables us to assign a confidence threshold to each UI type, balancing recall and precision (shown as orange points).}
  \Description{13 precision-recall curves (PR curves), each illustrating the model's detection performance for one UI type. Each PR curve is a line chart comparing the trade offs between precision and recall when applying different confidence thresholds for UI detection. Each PR curve also includes a point for our choice of confidence threshold, which err on the side of higher recall. Toggle (unselected) seems to have the highest performance, in which the precision remains close to 100\% as recall increases, with a small sudden drop at the end. Checkbox (selected) seems to have the lowest performance, which has a recall below 50\% for all confidence thresholds. The lines for Checkbox (selected), Dialog, Slider, and Toggle (selected) fluctuate more than the others, indicating that they have fewer data points.}
  \label{fig:PR_curves}
\end{figure}

Object detection models often return multiple bounding boxes for a given underlying target with different confidences. We use two post-processing methods to filter out duplicate detections: (i) Non-Max Suppression picks the most confident detection from overlapping detections \cite{neubeck2006efficient}. (ii) Different confidence thresholds can be applied on each UI type to remove less certain detections. We picked confidence thresholds per UI type that balance recall and precision for each (Figure ~\ref{fig:PR_curves}).

\subsection{Training and Testing Data}
The training and testing datasets include 13 classes (Table \ref{tab:average_precision}) derived from the UI types in annotation: Checkbox (Selected), Checkbox (Unselected), Container, Dialog, Icon, Picture, Page Control, Segmented Control, Slider, Text, Text Field, Toggle (Selected), and Toggle (Unselected). Based on initial experiments with UI detection models, we split the Checkbox and Toggle into ``Selected'' and ``Unselected'' subtypes, and remapped the Tab Bar Item into Icon and Text.

\textbf{Checkbox and Toggle:} The visual appearance of ``Selected'' and ``Unselected'' Checkboxes are quite different (same for Toggles), and splitting them improved the overall detection accuracy for each class. Directly detecting selection state also means we do not need a separate classification model to determine the selection state of a detected Checkbox or Toggle, which reduces on-device memory usage and inference time.

\textbf{Tab Bar Item}: Initial experiments revealed that the standalone Tab Bar Item class had good recall but low precision. This outcome was because our initial model tended to detect most UI elements at the bottom of screen as Tab Bar Items, including Icons and Text at the bottom of screen of apps that did not have Tab Bars at all. Therefore, we decided to split the Tab Bar Item annotations into Text and/or Icon classes. Later, we will describe heuristics that we introduced to group rows of Text and Icons near the bottom of the screen into Tab Bar Items contained within a Tab Bar.

Regarding our training/testing data split, a simple random-split cannot evaluate the model generalizability, as screens in the same app may have very similar visual appearances. To avoid this data leakage problem \cite{kaufman2012leakage}, we split the screens in the dataset by app. We also ensured the representation of app categories, screen count per app, and UI types are similar in the training and testing datasets. We ran several random splits on apps, and stopped when the split satisfied these requirements. The resulting split has \placeholder{72,635} screens in the training dataset and \placeholder{5,002} screens in the testing dataset.

\begin{table}[h!]
  \begin{center}
    \caption{\camerareadychange{Average Precision (AP) of Each UI Type in \placeholder{5,002} Testing Screens.}}
    \label{tab:average_precision}
    \begin{tabular}{lccr}\hline
    UI Type & AP (>0.5 IOU) & AP (Center) & Count\\ \hline
    Checkbox (Unselected) & 77.5\% & 79.1\% & 471 \\
    Checkbox (Selected) & 27.4\% & 27.4\% & 119 \\
    Container & 82.1\% & 83.0\% & 11528\\
    Dialog & 62.9\% & 63.3\% & 264 \\
    Icon & 79.7\% & 88.0\% & 21875 \\
    Page Control & 65.4\% & 87.3\% & 454 \\
    Picture & 72.0\% & 76.9\% & 9211 \\
    Segmented Control & 55.1\% & 57.6\% & 1036 \\
    Slider & 55.6\% & 63.0\% & 110 \\
    Text & 87.5\% & 91.7\% & 47045 \\
    Text Field & 79.2\% & 79.5\% & 1379 \\
    Toggle (Unselected) & 91.7\% & 91.7\% & 247 \\
    Toggle (Selected) & 90.5\% & 92.0\% & 131 \\ \hline
    \textbf{Mean} & \textbf{71.3\%} & \textbf{75.4\%} \\
    \textbf{Weighted Mean} & \textbf{87.5\%} & \textbf{83.2\%} \\  \hline
  \end{tabular}
  \end{center}
\end{table}

\subsection{Evaluation}

For each UI type, we evaluated our model performance using Average Precision (AP), a standard object detection evaluation metric. We choose a threshold of $>0.5$~IoU (Intersection over Union), commonly used in object detection challenges \cite{everingham2010pascal}, to match a detection with a ground truth UI element. On our testing dataset (\placeholder{5,002}~screenshots), our UI detection model achieved \placeholder{71.3\%}~mean AP. If we weigh our results by the frequency of each UI type in the dataset, the weighted mean AP is \placeholder{82.7\%}. Table ~\ref{tab:average_precision} shows the individual AP for each of the 13 UI types. Based on the precision-recall curves (PR curves) for individual UI types (as shown in Fig. \ref{fig:PR_curves}), we chose a different confidence threshold for each UI type. The use case of screen reader accessibility informed our choices. Note that the choices may differ for other accessibility services (\eg SwitchControl users may prefer higher precision over recall to reduce the number of elements to navigate when they can see elements that our model fails to detect).

\begin{figure}[t!]
  \centering
\includegraphics[width=\linewidth]{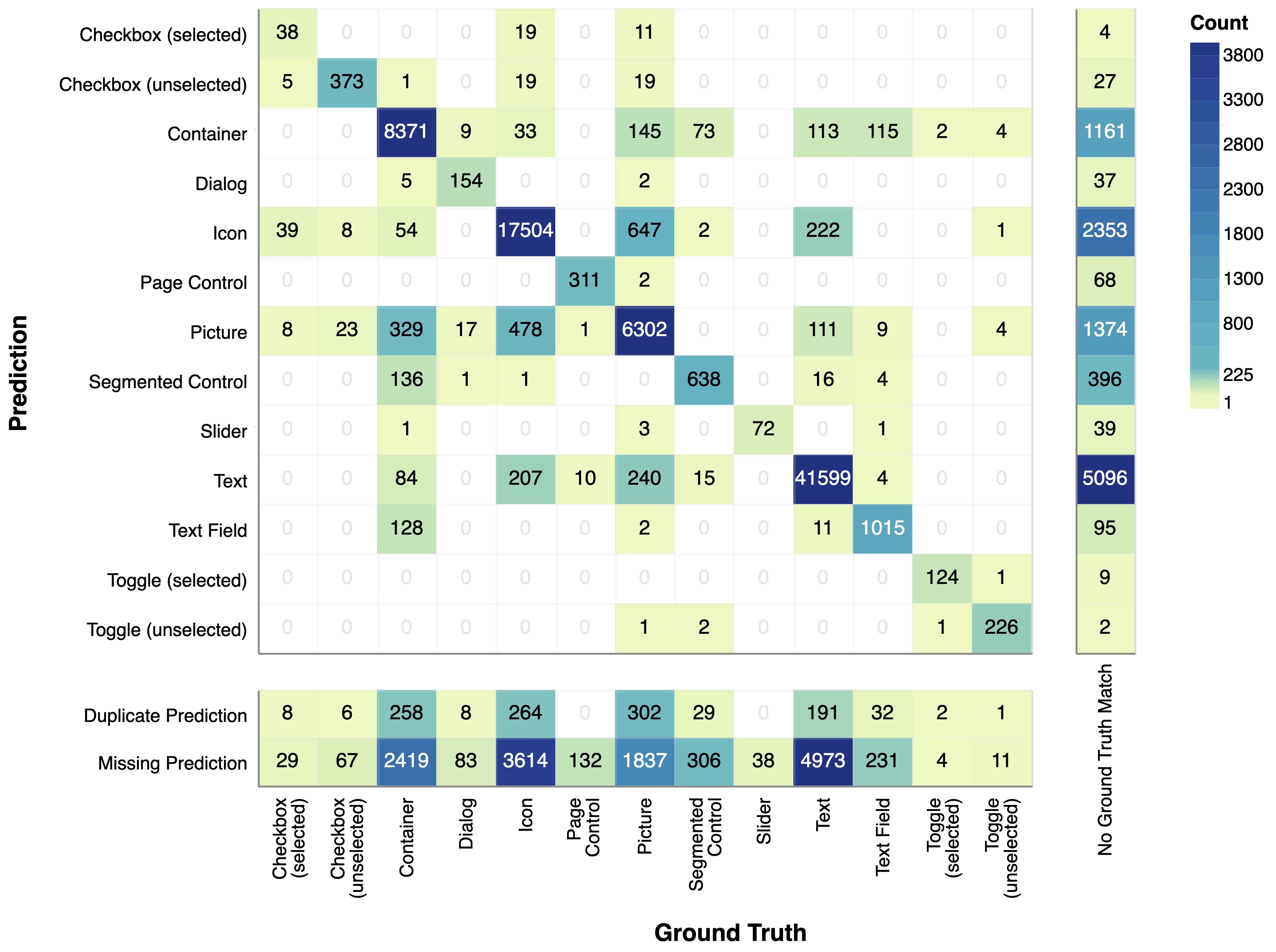}
    \caption{Confusion matrix for all UI types. Checkboxes (selected) were often confused with Icons or Pictures. Icons were sometimes confused with Pictures.}
    \label{fig:confusion_matrix}
    \Description{
    A confusion matrix of the model prediction results, showing the number of each predicted UI type for each ground truth UI type, enabling someone to examine how many predictions for one UI type were correct, and how many times the model confused it with a different particular UI type. It also shows the number of each predicted UI type that had no ground truth match, as well as the number of each ground truth UI type that had a duplicate prediction or missing prediction. 
    The model had the least confusion for Toggle (selected), in which it misclassified one selected Toggle as an unselected Toggle, missed only 4 predictions, and predicted 9 extraneous or non-existent Toggles. The model confused Pictures, Containers, and Icons with other UI element types more than others, classifying 647 Pictures as Icons, 478 Icons as Pictures, and 329 Containers as Pictures, and 240 Pictures as Text. Patterns in the confusion matrix also highlight the imbalanced representation of different UI types described in Section \ref{sec:datasetAnalysis}.
    }
\end{figure}

Our model achieved the lowest AP for the Checkbox (Selected). Its low frequency in the dataset may contribute to poor model performance; however, the Slider and Toggle also have a similar low frequency. To further understand how much the model misclassifies one UI type as another, we generated a confusion matrix \cite{santiago2018} as shown in Fig. ~\ref{fig:confusion_matrix}.
We found that the model often confuses Checkbox (Selected) with Icon and sometimes Picture. Checkboxes look visually similar to Icons, potentially explaining why our model tends to mis-classify them as Icons, which have much higher frequency in the dataset. Creating another classification model to distinguish Checkbox (Selected) from Icon detections has the potential to double its recall.

We also evaluated our model's detection performance with an evaluation metric specific for accessibility services: whether the center of a detection lies within its target UI element. With mobile screen readers, double-tapping a detection will pass a tap event to its center, which  activates the target. This evaluation metric is more relaxed than $>0.5$~IoU and increases the mean AP from \placeholder{71.3\%} to \placeholder{75.4\%}, as shown in column AP (Center) of Table ~\ref{tab:average_precision}. In some cases, the detection bounding box may not be tight enough to include all semantic information of a target UI element, but still enables users to manipulate the target. Thus, this metric may better capture whether the model enables people to use the manipulable UI elements.

\section{Improving the User Experience from UI Detection Results} \label{sec:improving}
Our model correctly detects and classifies most UI elements on each screen; however, it may still miss some UI elements and generate extra detections. Even when the detections are perfect, simply presenting them to screen readers will not provide an ideal user experience as the model does not provide comprehensive accessibility metadata (\eg UI content, state, clickability). We worked with 3 blind QA engineers and 2 senior accessibility engineers in our company to iteratively uncover and design the following \placeholder{6} improvements to provide a better screen reader experience. Over 5 months, the participating engineers tried new versions of the model on apps of their choice, and provided feedback on when it worked well and how it could be improved.

\subsection{Finding Missing UI Elements and Removing Extra Detections}
\begin{figure}[t!]
  \centering
  \includegraphics[width=\linewidth]{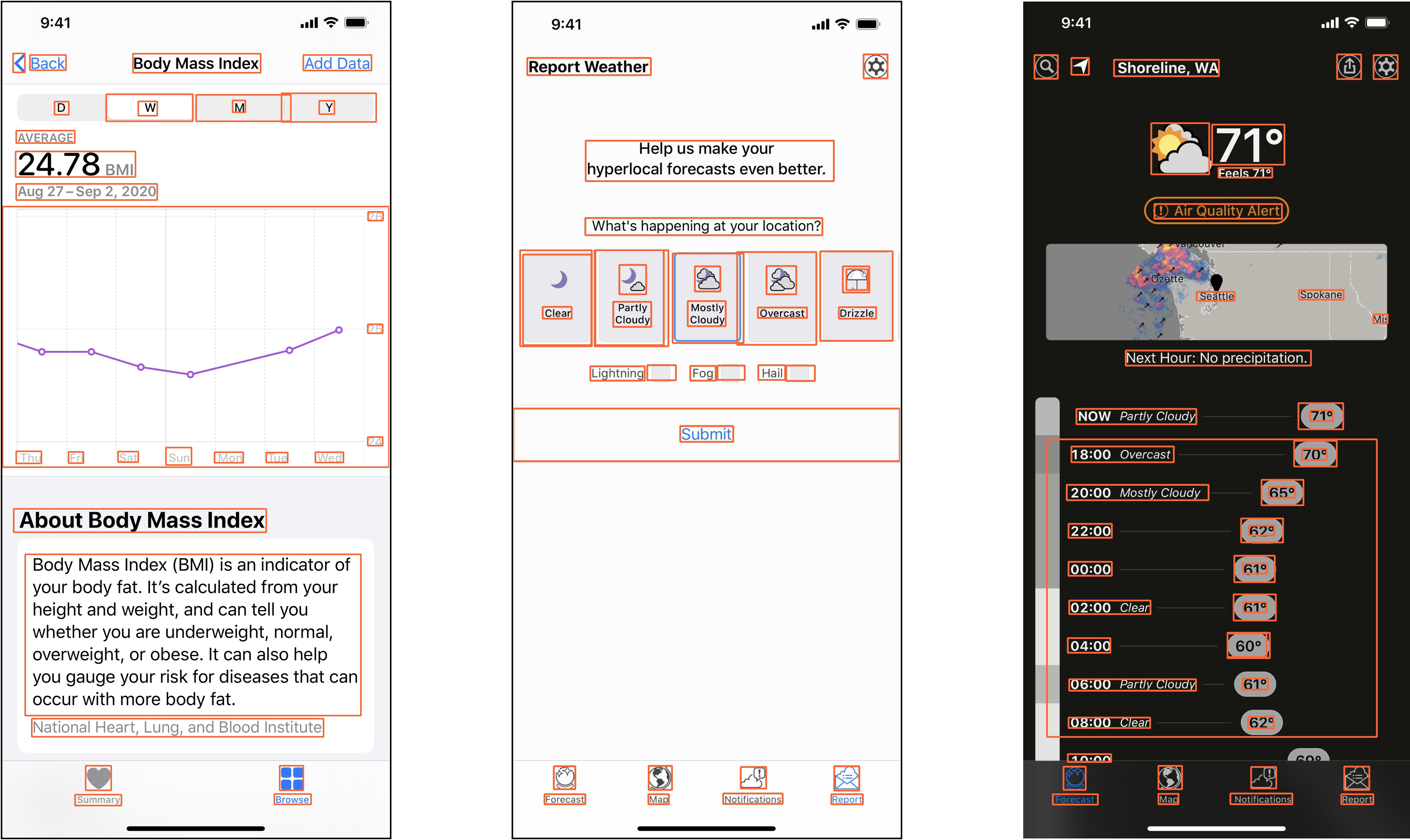}
  \caption{The raw UI detection results from our model would not always result in the best screen reader user experience. (Left) The model does not detect the leftmost Segmented Control. (Middle) The rightmost umbrella has two detections; Checkbox and Text are not grouped. (Right) The big Container detection at the bottom half may be removed. In all examples, Icon and Text on the tab bar should be grouped as Tab Buttons. The example screens are from Apple Health and Dark Sky.}
  \Description{Screenshots of our models raw UI detection results with red boxes around individual detected UI elements. The left screenshot has a red box missing around the leftmost segmented control. The right screenshot show picture elements with text elements below, each with an individual red box.} 
  \label{fig:raw_detection}
\end{figure}

Our UI detection model may miss some UI elements. As seen in Fig.~\ref{fig:raw_detection} Left, the model sometimes only detects a subset of Segmented Controls (SC). In 99.1\% of the screens in our dataset, every Text on the same row of SCs is contained by a SC. Based on this pattern, after we detect SCs in a screen, we find Text detections on the same row of detected SCs but not contained by any SC. Our heuristics update these Text elements as Segmented Controls. Another common missing UI element is Text, especially when it is small and/or on a complex background. To address this problem, we ran the iOS built-in OCR engine~\cite{appleOCR}, which is trained specifically to detect text and outperforms our model on this class. We include text detections from OCR when our model misses them (\ie OCR text has no overlap with existing detections).

Our UI detection model may also detect extra UI elements. For a UI element, our model sometimes generates multiple bounding boxes, each with a different but visually similar UI type (\ie Picture/Icon, Segmented Control/TextField/Container). For example, in Fig.~\ref{fig:raw_detection} (Middle), the rightmost "umbrella" has a box detected as Icon and another box detected as Picture. Therefore, we apply a customized Non-Max Suppression algorithm to remove duplicate detections within these visually similar UI types. There are also extra UI elements we should not remove (\eg a large Picture can contain a small Icon, a Container may contain a TextField), thus our algorithm only removes spatially similar detections when IoU~$>0.8$.

\subsection{Recognizing UI Content}
UI types Text, Icon, and Picture (comprising 83.3\% of UI elements in our dataset) contain rich information. Therefore, it is important to recognize and describe their contents to screen reader users. We leverage some iOS built-in features to recognize content in these UI detections. \textbf{For Text:} we use the iOS built-in OCR engine \cite{appleOCR}, which provides a tight bounding box and accurate text result with a reasonable latency (<~0.5s). \textbf{For Icon:} we leverage the Icon Recognition engine in iOS 13 (and later) VoiceOver to classify \placeholder{38} common icon types \cite{appleVoiceOver}. \textbf{For Picture:} we leverage the Image Descriptions feature in iOS 14's VoiceOver to generate full-sentence alternative text \cite{ios14preview}.

\begin{figure}[tb]
  \centering
  \includegraphics[width=\linewidth]{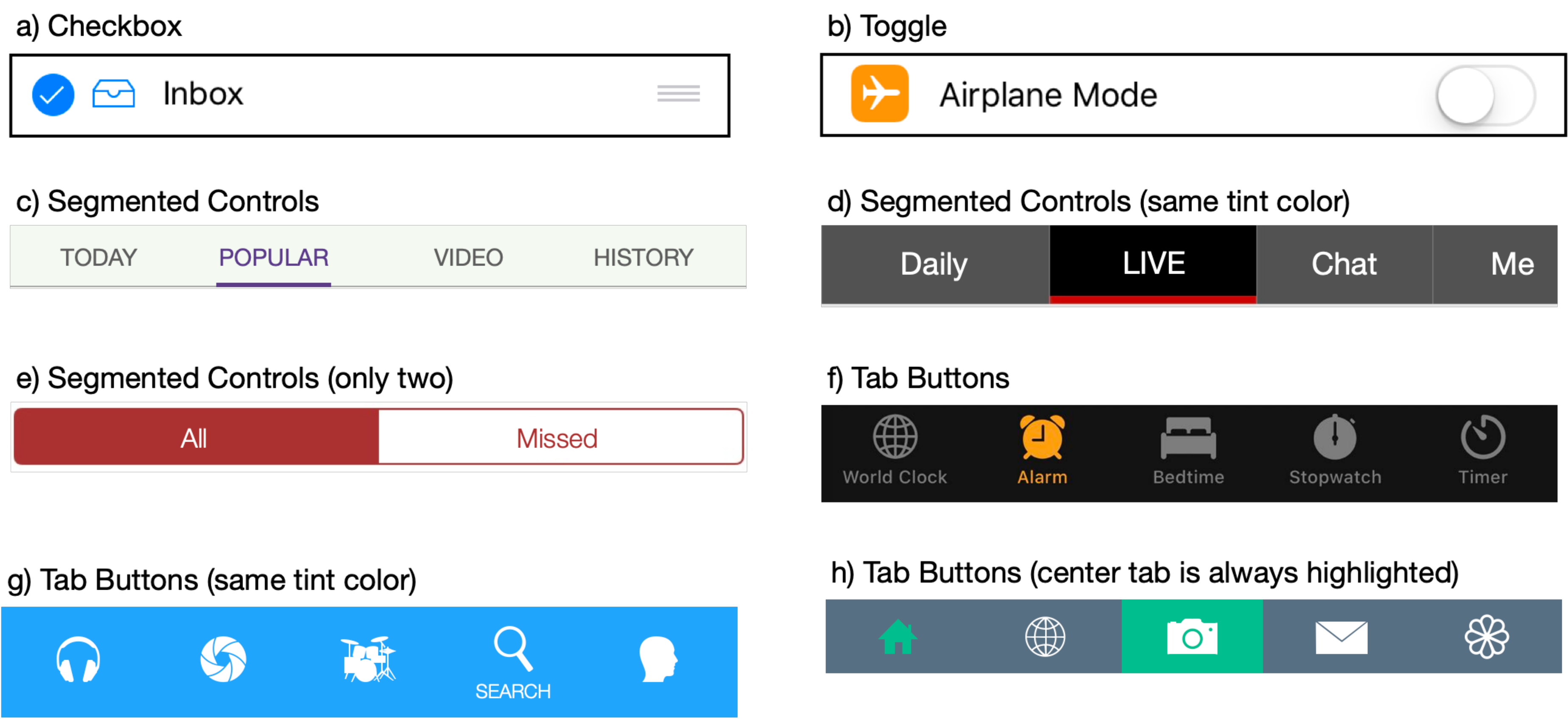}
  \caption{Examples of how selection state is visually indicated in different UI types. Differences in tint color is often a signal of selection for Segmented Controls and Tab Buttons. There are also other visual indicators for selection states.}
  \Description{The top row visualizes the selection state for Checkbox and Toggle elements (a and b) For a), the selected checkbox is indicated by a checkmark inside the checkbox. For b), the selection state is indicated by a toggle switch that moves to the right and changes color when selected. The second row visualizes the selection state for Segmented Controls. For c), the selection state is indicated by just changing the text color for the selected tab. For d) the selection state is indicated by changing the background color for the selected tab. For e), the selection state is indicated by changing both the background color and text color. In the the third row, we show examples of selection state for tab buttons. In f), the selected tab is indicated by changing the tint color of the selected icon and text. In g), the selection is indicated by adding a text label below the selected tab (where the unselected tabs have no text label). In h), the selected tab is indicated by changing the tint color, while the center tab is always highlighted with a different tint and background color.} 
  \label{fig:selection_state}
\end{figure}

\subsection{Determining UI Selection State}
Several UI types also have selection states, such as Toggle, Checkbox, Segmented Control, and Tab Button. With this metadata, VoiceOver can announce UIs with selection states (\eg "Airplane Mode, Off", "Alarm, Tab, Selected" in Fig. \ref{fig:selection_state}). For Toggle and Checkbox, our UI detection model includes selection states in detected UI types. For Segmented Control and Tab Button, we leverage their visual information to determine the selection states. The most common visual indicator is the tint color. As shown in Fig. \ref{fig:selection_state}.f, the selected Tab has tint color while other Tabs are not highlighted. Within each Tab, we extract the most frequent color as the background color, and the second most frequent color as the tint color. Among all Tab Buttons, we find the one with an outlier tint color \cite{colorCube}. Finally, we assign a "selected" state to that detection, and a "not selected" state to the remaining detections. Segmented Controls also frequently use tint color as a visual indicator. As shown in Fig. \ref{fig:selection_state}, custom UI designs may keep the same tint color and use other visual indicators (Fig. \ref{fig:selection_state}.c); therefore, we tried to address some common designs with additional heuristics, such as checking the bar at the bottom of the Segmented Control (Fig. \ref{fig:selection_state}.d), or finding the only Tab with text (Fig. \ref{fig:selection_state}.g). We evaluate our selection state heuristics on our testing dataset (\placeholder{5002} screens). On the 936 screens in which our approach detects Tab Buttons, we assign the correct selection state \placeholder{90.5\%} of the time. Among 337 screens in which our approach detects Segmented Controls, we assign the correct selection state on \placeholder{73.6\%} of screens.

\subsection{Determining UI Interactivity}
Knowing whether a UI element is interactive is important for screen reader users. With the correct accessibility metadata, a screen reader can announce actions supported by each UI element (\eg "Page X of Y, swipe up or down to adjust the value" for a page control; "Password, double tap to edit" for a text field; "Menu, button" for a clickable UI element like a button, web link, table cell, etc). Several UI types detected by our model (\eg Text Field, Page Control, Slider) already indicate their supported actions. On the other hand, Text, Picture, and Icon can be non-clickable in some cases (\eg Airplane Icon, "Inbox" Text in Fig. ~\ref{fig:selection_state}), but clickable in other cases (\eg Back Icon, "Add Data" Text in Fig. ~\ref{fig:raw_detection}). According to feedback from screen reader users, they can often infer the clickability of Text and Picture elements from alternative text (generated by OCR and Image Descriptions), but have a difficult time telling if an Icon is clickable from the Icon Recognition result (in addition, it is unavailable when the Icon is not in the 38 supported icon types). To predict Icon clickability, we trained a Gradient Boosted Regression Trees model \cite{boostedTrees} with several features (\ie location, size, Icon Recognition result). \camerareadychange{This model is trained with TuriCreate Boosted Trees Classifier toolkits \cite{boostedTrees}; it takes less than 10ms for inference and uses 10MB of memory (CoreML model).} We only mark an Icon as clickable when our model is very confident, as screen reader users prefer higher precision over recall: if a clickable Icon is predicted as non-clickable, the user may still try to activate it; if a non-clickable Icon is predicted as clickable, the user will hear "Button" in screen reader announcements, and will be confused if the activation does not do anything. Therefore, we pick a threshold for our model to keep \placeholder{90.0\%} precision, which allows \placeholder{73.6\%} recall. Additionally, there are more UI interactivities worth investigating in future work, such as detecting the enabled/disabled state of UI elements.

\begin{figure*}
  \centering
  \includegraphics[width=\textwidth]{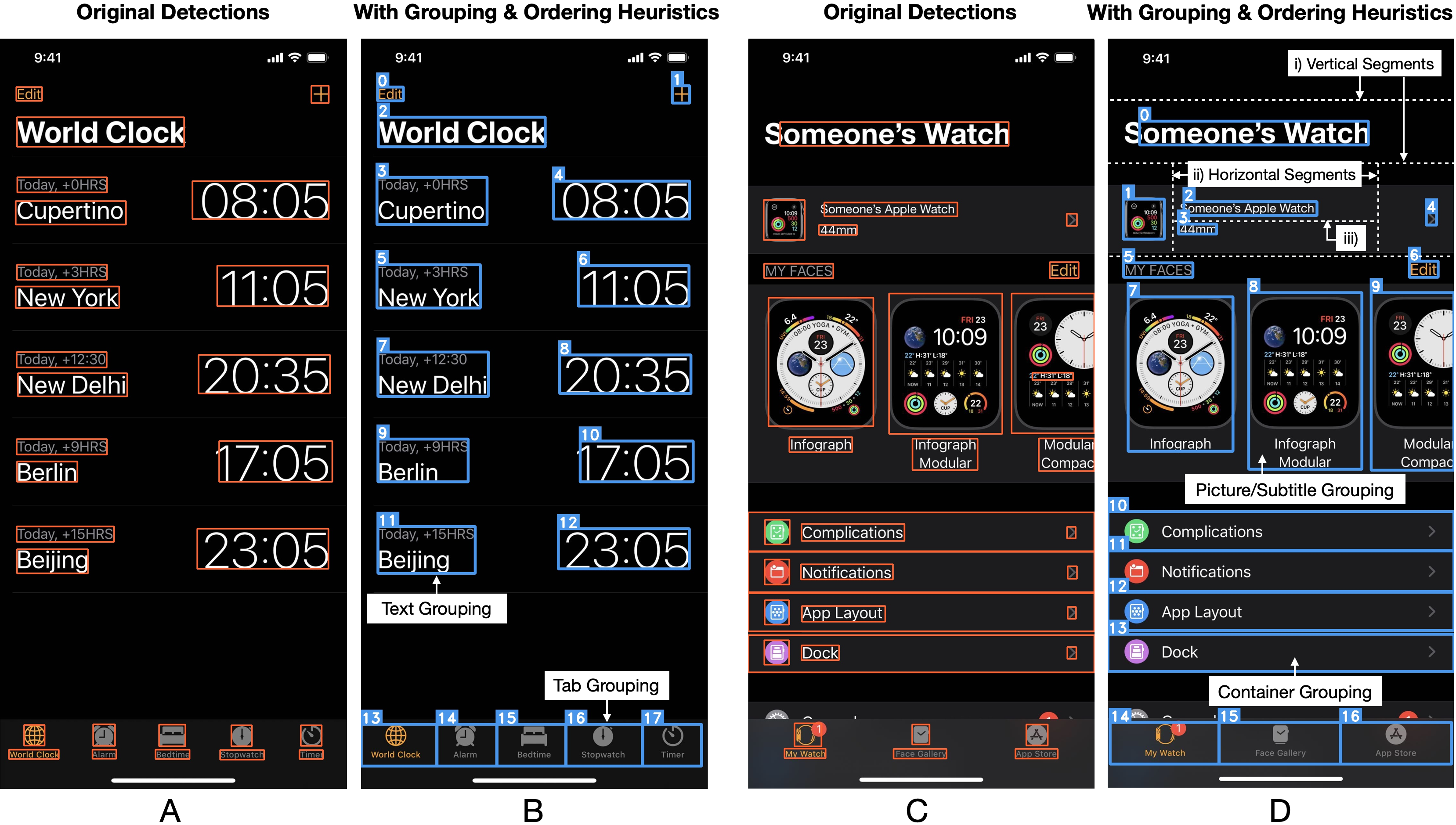}
  \caption{Example screens to demonstrate our grouping and ordering heuristics: {\em (A, C)} UI detections from our model in screens of Apple Clock and Apple Watch apps, {\em (B)} the output demonstrating text and tab button grouping, and {\em (D)} the output demonstrating picture/subtitle and container grouping. Our ordering heuristics infer the navigation order (shown on the top-left of each box), which recursively divides the screen into vertical and horizontal regions (visualized examples with dashed line).}
  \Description{4 screenshots from Apple Clock app and Apple Watch app. Screenshots A and C show red boxes around each individual detected UI element without any grouping or ordering. Screenshot B shows blue boxes around each UI element where some have been grouped, specifically text with text below it, and nd tab icons with their labels have been grouped with a blue border. Numerical annotations marking the navigation order are shown beside each element going in order down the screenshot. Screenshot D also shows blue boxes around UI elements, some of which have been grouped. UI elements inside of a container row (i.e., an icon, text, and right arrow), and Picture with text below have each been grouped together with a blue border. Numerical annotations are also shown beside each element to mark the navigation order.}
  \label{fig:grouping_ordering_heuristics}
\end{figure*}

\subsection{Grouping Elements for Efficient Navigation}
Grouping related UI elements makes navigation more efficient for screen readers and other accessibility services \cite{switchaccess,switchcontrol}. Fig. ~\ref{fig:grouping_ordering_heuristics} shows the original detections from our UI element detection model; navigating through a large amount of detections would take a long time for screen reader users to understand the screen contents. To reduce the number of UI elements in navigation, we group related elements together. For grouped UI elements, we provide access to clickable sub-elements through custom actions (supported by mobile screen readers). The user can also read each non-clickable sub-element by adjusting the speech granularity in the rotor \cite{rotor}. This approach still has limitations: when our model makes mistakes in inferring clickability, users may lose access to some clickable sub-elements. However, screen reader users noted that the increased efficiency from grouping was worth the tradeoffs of losing access to some elements.

We developed multiple heuristics that group UI detections based on their UI types, sizes, and spatial relationships on the screen. Fig. ~\ref{fig:grouping_ordering_heuristics} displays the results of our grouping heuristics on example screens. We provide the concatenated text of sub-elements as alternative text for each grouped element.

\begin{itemize}
    \item \textit{Text Grouping} (Fig. ~\ref{fig:grouping_ordering_heuristics}.B): We group a Text $T_{1}$ with a Text below $T_{2}$ if they satisfy: 1) they have x-overlap, and 2) the y-distance between the two texts should be less than a threshold -- we choose $min(T_{1}.height, T_{2}.height)$. We do this repetitively to group multi-line text.
    \item \textit{Tab Grouping} (Fig. ~\ref{fig:grouping_ordering_heuristics}.B): We group Icon and Text detections into Tab Buttons by 1) finding the bottom-most detection $B$; 2) determining if a Text or Icon detection $D$ is in the tab bar if $D$ is in the bottom 20\% of the screen, and $B.bottom$ - $D.top$ is within a threshold (0.08 * screen height) so that $D$ is not much higher than $B$; 3) grouping Icon and Text detections that have x-overlap together, and determined to be in the tab bar, as one Tab Button.
    \item \textit{Container Grouping} (Fig. ~\ref{fig:grouping_ordering_heuristics}.D): We group a Container detection with all detections inside it as one element. 
    \item \textit{Picture Subtitle Grouping} (Fig. ~\ref{fig:grouping_ordering_heuristics}.D): We group Text $T_{1}$ as a descriptive subtitle with Picture $P$ if they satisfy: 1) they have 50\%+ x-overlap with $P$, 2) $T_{1}$ has a y-distance to $P$ less than a threshold (we chose 0.03 * screen height), and 3) $T_{1}$ is closer to $P$ than to any other detection below $T_{1}$. We additionally group a second line of Text $T_{2}$ below $T_{1}$ if it satisfies the same conditions in relation to $T_{1}$.
        
\end{itemize}

To evaluate these heuristics, we ran our grouping heuristics on a subset of collected screens (n=300, randomly picked). Overall, our grouping heuristics reduced the number of UI elements to navigate by 48.5\% (Without Grouping - Mean: 21.83, Std: 12.8, With Grouping - Mean: 12.1, Std: 7.6). Two researchers (authors of this paper) manually tabulated grouping errors across these screens using a rubric which contains 4 grouping heuristic categories (\ie text grouping, container, picture subtitle, tab), and "Other" for additional errors. Grouping errors are classified as "Should have grouped" if two or more elements should have been grouped together, or "Incorrectly grouped" if two or more unrelated elements were grouped. As there may be multiple groupings for some elements, we did not count a grouping or missed grouping as an error if it was ambiguous. We acknowledge that human observation is not perfect, and there might be bias in doing evaluation by the authors. To mitigate these concerns, the researchers first independently evaluated 10\% of screens and then discussed the errors found until reaching a sufficient level of agreement. After reaching consensus, each researcher examined half of the screens, and found very similar numbers of grouping errors across independent sets of screens. Among 1568 groups made, 88 (5.6\%) were "Incorrectly grouped". In addition, we found 99 "Should have grouped" errors. There were 0.62 errors per screen (Min=0, Max=8, Std=1.1).

The largest set of grouping errors were from our \textit{text grouping} heuristic (25.7\%), followed by \textit{picture subtitle} (16\%), \textit{tab button} (8.6\%) and \textit{container} (7.5\%). 42\% of the grouping errors did not fall into one of our heuristic categories: the majority of errors (48 errors) consisted of missed grouping text with contained icons, remaining OCR-detected text inside icons, and grouping of unrelated text. While evaluating grouping errors, we also found opportunities to further improve the user experience (\eg group text and text field below it), which may lead to new heuristics.

\camerareadychange{While our heuristics in general performed well, we note that these heuristics are likely to be more successful when applied to iOS app screens due to the influence of native user interface widgets and design guidelines (\ie Human Interface Guidelines). For example, our Tab Grouping heuristic likely works well because iOS provides a standard widget, which a significant amount of apps use, to display tabs at the bottom of an app screen (i.e., UITabBarController). As such, our heuristics may need to be adapted for other platforms (\eg Android) if the platform's widget designs are quite different from iOS designs.} 

\subsection{Inferring Navigation Order}
Mobile screen readers allow swiping to navigate all UI elements on a screen in a logical order. Therefore, we need to provide our detected UI elements to screen readers in a reasonable navigation order. After applying our grouping heuristics, we determine the navigation order with a widely used OCR XY-cut page segmentation algorithm~\cite{nagy1984hierarchical, ha1995recursive}, which sorts text/picture blocks of a document in human reading order. We apply the algorithm to our problem, as demonstrated in Fig. ~\ref{fig:grouping_ordering_heuristics}.D: (i) We build vertical segments by determining Y coordinates where we can draw a horizontal line across the screen without overlapping any element. (ii) Within each vertical segment, we build horizontal segments by locating X coordinates where we can draw a vertical line down the segment without overlapping any element. (iii) We recursively call this algorithm on all horizontal segments to further create vertical segments and so forth. When we can no longer subdivide a segment, we order the elements inside by top-to-bottom. If there is a tie, we order them left-to-right. Within grouped elements, we also apply the XY-cut algorithm to order sub-elements. Fig. ~\ref{fig:grouping_ordering_heuristics} shows a number on the top-left of each element to indicate the navigation order we provide to screen readers. In Fig. ~\ref{fig:grouping_ordering_heuristics}.B, a screen reader user would hear "Edit", "Add", "World Clock", and "Today, plus 0 HRS, Cupertino" for the first four elements.

To evaluate our ordering heuristics, we ran them on a subset of collected screens (n=380, randomly picked). Ten accessibility experts in our company annotated the ground truth navigation order. For each screen, we compared our heuristics output and the ground truth navigation order with a metric often seen in list sorting: the minimum insertion steps to make our output the same as ground truth (in one step, an element can be inserted to any other position in list). Our heuristics output perfectly matched the ground truth for 280 screens (73.7\%), while 345 screens (90.8\%) had less than 1 error (\ie 1 insertion step) for every 10 elements. On average, there were 18.7 elements per screen (min=1, max=92, std=13.8); our heuristics produced a mean of 0.67 errors per screen (min=0, max=17, std=1.79).

To understand where our ordering heuristics could be improved, we examined 35 screens with more than 1 error for every 10 elements. On 12 screens, we found the ordering was ambiguous (\ie there were multiple reasonable orderings). For the remaining screens, 19 of them would have a better navigation order by improving our grouping heuristics: (i) our heuristics grouped some text with pictures incorrectly, or our \textit{picture subtitle} grouping heuristic did not group some text as a subtitle which caused ordering errors (13 screens); (ii) our \textit{text grouping} heuristics grouped some text incorrectly or should have grouped some text which caused ordering errors (9 screens); (iii) our heuristics detected some icons inside of text blocks but did not group them with the text, so the icons were ordered after the text (6 screens).

\section{User Studies}

To understand how our approach impacts the user experience of screen readers, we conducted a user study with \placeholder{9} \camerareadychange{people} with visual impairments \camerareadychange{(3 Female, 6 Male). They are all regular iOS VoiceOver users with an average of 8.9 years (min = 6, max = 11) of VoiceOver experience.} \camerareadychange{The average age of participants was 44.2 years (min = 28, max = 65).} In the studies, we asked participants to use apps with regular VoiceOver and \camerareadychange{Screen Recognition} (our approach) and compare their experiences. There were two types of studies: {\em (i)} Interview studies: through video conferencing with 4 screen reader users (1F, 3M) who are accessibility QA engineers in our company, and {\em (ii)} Email studies: a remote unsupervised usability study through emails with \placeholder{5} screen reader users (2F, 3M) who volunteered to participate, \placeholder{which provides more time flexibility than interviews}. To reduce bias, we excluded the QA engineers who provided suggestions in Section \ref{sec:improving} from our studies.

\subsection{Procedure}

Both studies used the same instructions and questions. We asked participants to select 3 iOS apps that they found difficult or impossible to use with VoiceOver. If participants could not think of such apps, we offered a list of suggestions. We asked the participants to spend 10 minutes using each app, first with regular VoiceOver and then with \camerareadychange{Screen Recognition}. \camerareadychange{During the study, participants were always aware of the current condition they were testing. We did not counterbalance the order of conditions, as all participants were experienced VoiceOver users and most apps in the study were suggested by participants (they already tried these apps with regular VoiceOver before).} We collected feedback from participants about the differences they experienced between regular \camerareadychange{VoiceOver} and \camerareadychange{Screen Recognition}, benefits from \camerareadychange{Screen Recognition}, and challenges encountered using \camerareadychange{Screen Recognition}.

Due to the COVID-19 pandemic, we conducted all user studies remotely. In the Interview studies, we sent study preparation through email in advance and feedback was collected through a conference call. In the Email studies, we sent and collected all study instructions and feedback via email.

\begin{figure}[t]
  \centering
  \includegraphics[width=\linewidth]{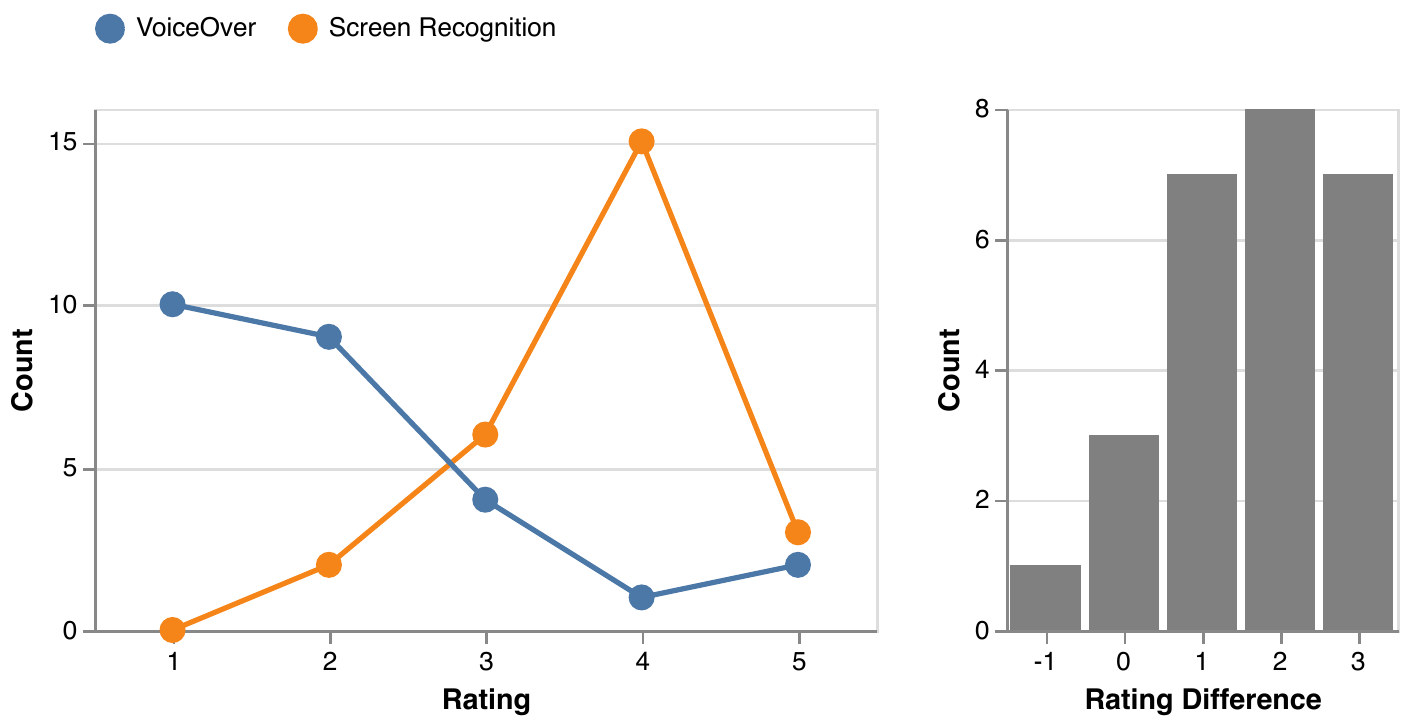}
  \caption{\camerareadychange{(Left) A histogram of usability ratings for apps using regular VoiceOver and Screen Recognition. (Right) A histogram of the rating differences within individual apps between the two systems, as the rating for Screen Recognition minus the rating for regular VoiceOver.}}
  \Description{Two bar charts comparing user ratings of using regular VoiceOver versus Screen Recognition for apps. On average, ratings for Screen Recognition are higher than for regular VoiceOver. The first bar chart shows how often each system received a particular rating, with the following breakdown: a rating of 1: applied to VoiceOver for 10 apps, and to Screen Recognition for 0 apps. Rating 2: VoiceOver 9, Screen Recognition 2. Rating 3: VoiceOver 4, Screen Recognition 6. Rating 4: VoiceOver 1, Screen Recognition 15. Rating 5: VoiceOver 3, Screen Recognition 3. The second bar chart shows how often Screen Recognition was rated a particular number of units higher than regular VoiceOver for an individual app, with the following breakdown: A difference between Screen Recognition’s rating and regular VoiceOver’s rating of -1: 1 app. A difference of 0: 3 apps. Difference of 1: 7 apps. Difference of 2: 8 apps. Difference of 3: 7 apps.}
  \label{fig:user_study_histogram}
\end{figure}

\subsection{Results}
Every participant chose apps to use during the study that they had previously found difficult to use with VoiceOver. Participants used 22 different apps, only 2 of which came from our suggested list. Only 2 apps were explored by multiple participants: one from our suggested list (chosen by 4 participants), and one not on the list (chosen by 2 participants). Note that we do not share specific app names because we want to be respectful and not single out individual developers. Participants chose a mix of apps with varying usability levels for VoiceOver and rated them from 1 (not usable at all) to 5 (completely usable).  Most participants chose 3 apps that they already knew to be inaccessible, but 3 participants also decided to include an app that they knew to be accessible which provided useful insights. All apps that participants chose were available on the iOS App Store as of August 2020. Some apps were built by major tech companies, but most were made by smaller developers or non-tech companies, such as media companies.

Overall, usability ratings for apps using regular VoiceOver averaged 2.08~(SD~1.20), whereas ratings for apps using \camerareadychange{Screen Recognition} averaged 3.73~(SD~0.78). This difference is significant according to a \camerareadychange{Wilcoxon signed-rank test (p~<~0.00004)}. \camerareadychange{Figure \ref{fig:user_study_histogram} shows a histogram of usability ratings, and a histogram of rating differences within individual apps between the two systems.} It was clear to participants that the researchers created \camerareadychange{Screen Recognition}, so these results may reflect some bias \cite{likertbias, dell2012yours}, but still show quite a meaningful difference between the two systems, especially on the inaccessible apps that participants picked. In 4 cases, participants gave \camerareadychange{Screen Recognition} an identical or worse usability rating than regular VoiceOver. Two of these cases were for apps that the participant rated very usable (5) with regular VoiceOver, and then still rated \camerareadychange{Screen Recognition} as usable with either a 4 or 5. In the other two cases, both experiences received the same low rating (2 and 3 respectively). Based on participants' feedback and our own use, we believe these latter ratings reflect that these apps are difficult to use in general, even for sighted individuals.

For apps that participants rated unusable with regular VoiceOver, all noted that these apps exposed user interface elements without labels, and that it was frustrating to know that a control existed but not know its purpose. When exploring the same apps with \camerareadychange{Screen Recognition}, all participants remarked that they became aware of user interface elements that they did not realize existed because they were not exposed in the regular VoiceOver interface. One participant noted that in one app a slider became available, which they found surprising because these elements are often unavailable or unusable.  In another case, a participant said:

\quotateblock{QA2}{I am so excited I just NEED to share. Last night I downloaded [app name removed]...which has no AX elements and is completely unusable. So I decided to enable [\camerareadychange{Screen Recognition}]. Guess who has a new high score? I am in AWE! This is incredible. It is the first mainstream game I’ve ever been able to play on iOS besides Trivia Crack.}

Three participants chose to try apps that they knew to be accessible. In these cases, the participants still compared \camerareadychange{Screen Recognition} favorably to regular VoiceOver, remarking that \camerareadychange{Screen Recognition} exposed certain features in a different and potentially useful way. For example, one participant noted that the screen layout was revealed to them in a way that they often missed, saying:

\quotateblock{P1}{It was fun to discover the contents of the profile pictures displayed next to the person name and understand the layout of the app. Obviously, when making the app accessible, we oftentimes lose the understanding of the screen layout. With [\camerareadychange{Screen Recognition}] we now have the ability to understand the app spatially in addition to contextually.}

The only \camerareadychange{Screen Recognition} experience rated as less usable than the regular VoiceOver experience occurred with a social media app. The participant that introduced this app noted that it was very accessible and gave it a usability score of 5 with regular VoiceOver, whereas they gave \camerareadychange{Screen Recognition} a usability score of 4. The primary reason for this difference was the grouping choices from app developers, which made navigating through many tweets quick and focused on only the essential content. The participant noted that navigating this interface with \camerareadychange{Screen Recognition} was much slower, which includes a variety of low priority content (\eg number of likes, number of retweets, hash tags, etc.) that the app interface with regular VoiceOver omitted. However, this participant noted that they saw value in the \camerareadychange{Screen Recognition} interface in certain situations, because, although slow, it did allow access to certain content that wasn't available otherwise. Being able to switch between the two modes enabled this participant to drill down into the interface content when necessary, while also enjoy a streamlined experience the rest of the time.

This example illustrates an important result: Even with \camerareadychange{Screen Recognition}, the most accessible experiences are still created by app developers who understand how the content can best be conveyed to meet users' needs. Our system may not have the higher-level understanding of app interfaces to realize that the tweet list should be streamlined, while app developers can better understand the design and implement accessibility experience for users.

\camerareadychange{Screen Recognition} has room for improvement. Two participants noted that while many controls became apparent with \camerareadychange{Screen Recognition}, the labels on these controls sometimes required interpretation.  For example, an icon labeled ``down arrow'' might cause an item to move down in a list, but did not expose this higher level semantic. In these cases, the participants remarked that they were able to figure out the user interface with some experimentation. Another limitation of \camerareadychange{Screen Recognition} is that it cannot expose features that do not have a clear visual affordance. For example, a participant discovered that with \camerareadychange{Screen Recognition}, it was impossible to turn the pages in a book reading app, even though \camerareadychange{Screen Recognition} made the previously unreadable book text available. In this app, the page turning was actuated by a swipe from the screen edge, which had no visual affordance for \camerareadychange{Screen Recognition} to recognize. A challenging problem for future work could be to better understand the semantics of the visual interface. Finally, one participant noted that they occasionally experienced lag with \camerareadychange{Screen Recognition} on especially complex screens, which helps justify our intentions to later explore optimizations for our system.
\section{Discussion \& Future Work}

Despite persistent effort in developer tools, education, and policy, accessibility remains a challenge across platforms because many apps are not created with sufficient semantic metadata to allow accessibility services to work as expected. In this paper, we introduced an approach that instead generates accessibility metadata from the pixels of the visual interfaces, which opens a number of opportunities for future work.

A number of next steps follow directly from the approach and results presented in this paper. A clear next step is to keep improving our model and our heuristics to make them more accurate. Our current approach generates metadata from scratch over and over again; future work may explore how to allow this metadata to persist across invocations of an app or even across devices. Another next step is to apply our approach to other mobile platforms (\eg Android) and also beyond the mobile context, such as for desktop or Web interfaces, which are similarly inaccessible due to a lack of appropriate semantics provided by developers. Applying our approach to the desktop or Web would introduce new challenges, as our current method takes advantage of the size and complexity constraints of mobile apps. 

In the work presented here, we used only pixel information to create the semantic metadata needed for accessibility services. We did not leverage the semantic data inside the app view hierarchy, which accessibility services use \cite{appleVoiceOver, googleTalkback}. While combining information from our approach and view hierarchy could reconstruct a more complete set of accessibility metadata, we found that implementing this idea presents engineering challenges for effectively extracting and merging the two representations on-the-fly. Currently, users must enter a separate mode to access the automatic metadata; as we explore combined modes, we suspect we will need to even more carefully consider how to convey the inferred metadata to users. Furthermore, questions remain about how to resolve incompatible differences between the representations, especially in regard to higher-level concepts, such as the grouping and ordering. Instead of post hoc merging, an interesting opportunity for future work could be to build a model that leverages the structured information in the view hierarchy in addition to the pixels.

Our model and heuristics have multiple other use cases to explore. We demonstrated how we can apply them to improve accessibility on users' devices in real time. Another opportunity is to incorporate the model into developer tools to help developers create more accessible apps from the start. Current accessibility evaluation tools mainly depend on the (potentially incomplete) metadata developers already provide, limiting the tools' access to information to inform their accessibility suggestions. Our approach for understanding UIs from pixels could enable developer tools to identify more accessibility issues and automatically suggest fixes. Finally, while we have focused on accessibility use cases, the accessibility APIs also form the foundation of most on-device automation \cite{sugalite}, and so we expect our approach to find utility in many other use cases.

The longevity of our model is uncertain. Visual app designs change over time, both in terms of aesthetic qualities and in relation to their target devices. While people are generally able to adapt to such visual changes, our model may need to re-trained. It would be ideal to create a more universal model that is resilient to such changes or could even update itself while visiting new apps. While the uncertainty remains, our dataset, as we update it over time, will provide a window into how these visual changes occur and how they affect automatic methods operating on them.

Finally, our approach could be seen as reactive to the way most app development currently prioritizes visual design, instead of approaching the problem from an accessibility first perspective. We hope that by automating some aspects of accessibility metadata generation, especially through future enhancements to developer tools, our approach could help scale basic accessibility, freeing researchers, developers, and accessibility professionals from playing catch up, and instead enable them to work toward more innovative and comprehensive accessible experiences for all.

\section{Conclusion}
We have presented an approach to automatically create accessibility metadata for mobile apps from their pixels. Our technical evaluation and user evaluation demonstrate that this approach is promising, and can often make inaccessible apps more accessible. Our work illustrates a new approach for solving a long-standing accessibility problem, which has implications across a number of different platforms and services. Going forward, we would like to use the auto-generated accessibility metadata to not only directly impact accessibility features, but also help developers make their apps more accessible from the start.

\begin{acks}
\camerareadychange{We thank our colleagues at Apple, in particular: Cynthia L. Bennett, Shantanu Chhabra, Dominik Moritz, and Jay Pillai for their contributions.}
\end{acks}

\bibliographystyle{ACM-Reference-Format}
\bibliography{references}

%%% -*-BibTeX-*-
%%% Do NOT edit. File created by BibTeX with style
%%% ACM-Reference-Format-Journals [18-Jan-2012].

\begin{thebibliography}{81}

%%% ====================================================================
%%% NOTE TO THE USER: you can override these defaults by providing
%%% customized versions of any of these macros before the \bibliography
%%% command.  Each of them MUST provide its own final punctuation,
%%% except for \shownote{}, \showDOI{}, and \showURL{}.  The latter two
%%% do not use final punctuation, in order to avoid confusing it with
%%% the Web address.
%%%
%%% To suppress output of a particular field, define its macro to expand
%%% to an empty string, or better, \unskip, like this:
%%%
%%% \newcommand{\showDOI}[1]{\unskip}   % LaTeX syntax
%%%
%%% \def \showDOI #1{\unskip}           % plain TeX syntax
%%%
%%% ====================================================================

\ifx \showCODEN    \undefined \def \showCODEN     #1{\unskip}     \fi
\ifx \showDOI      \undefined \def \showDOI       #1{#1}\fi
\ifx \showISBNx    \undefined \def \showISBNx     #1{\unskip}     \fi
\ifx \showISBNxiii \undefined \def \showISBNxiii  #1{\unskip}     \fi
\ifx \showISSN     \undefined \def \showISSN      #1{\unskip}     \fi
\ifx \showLCCN     \undefined \def \showLCCN      #1{\unskip}     \fi
\ifx \shownote     \undefined \def \shownote      #1{#1}          \fi
\ifx \showarticletitle \undefined \def \showarticletitle #1{#1}   \fi
\ifx \showURL      \undefined \def \showURL       {\relax}        \fi
% The following commands are used for tagged output and should be
% invisible to TeX
\providecommand\bibfield[2]{#2}
\providecommand\bibinfo[2]{#2}
\providecommand\natexlab[1]{#1}
\providecommand\showeprint[2][]{arXiv:#2}

\bibitem[\protect\citeauthoryear{Apple}{Apple}{2020a}]%
        {rotor}
\bibfield{author}{\bibinfo{person}{Apple}.} \bibinfo{year}{2020}\natexlab{a}.
\newblock \bibinfo{title}{About the VoiceOver rotor on iPhone, iPad, and iPod
  touch}.
\newblock
\newblock
\urldef\tempurl%
\url{https://support.apple.com/en-us/HT204783/}
\showURL{%
\tempurl}


\bibitem[\protect\citeauthoryear{Apple}{Apple}{2020b}]%
        {iosax}
\bibfield{author}{\bibinfo{person}{Apple}.} \bibinfo{year}{2020}\natexlab{b}.
\newblock \bibinfo{title}{Accessibility for Developers}.
\newblock
\newblock
\newblock
\shownote{https://developer.apple.com/accessibility/.}


\bibitem[\protect\citeauthoryear{Apple}{Apple}{2020c}]%
        {boostedTrees}
\bibfield{author}{\bibinfo{person}{Apple}.} \bibinfo{year}{2020}\natexlab{c}.
\newblock \bibinfo{title}{Boosted Trees Classifier}.
\newblock
\newblock
\urldef\tempurl%
\url{https://apple.github.io/turicreate/docs/userguide/supervised-learning/boosted_trees_classifier.html}
\showURL{%
\tempurl}


\bibitem[\protect\citeauthoryear{Apple}{Apple}{2020d}]%
        {ios14preview}
\bibfield{author}{\bibinfo{person}{Apple}.} \bibinfo{year}{2020}\natexlab{d}.
\newblock \bibinfo{title}{iOS 14 Preview - Features}.
\newblock
\newblock
\newblock
\shownote{https://www.apple.com/ios/ios-14-preview/features/.}


\bibitem[\protect\citeauthoryear{Apple}{Apple}{2020e}]%
        {appleOCR}
\bibfield{author}{\bibinfo{person}{Apple}.} \bibinfo{year}{2020}\natexlab{e}.
\newblock \bibinfo{title}{Recognizing Text in Images}.
\newblock
\newblock
\urldef\tempurl%
\url{https://developer.apple.com/documentation/vision/recognizing_text_in_images/}
\showURL{%
\tempurl}


\bibitem[\protect\citeauthoryear{Apple}{Apple}{2020f}]%
        {turicreate}
\bibfield{author}{\bibinfo{person}{Apple}.} \bibinfo{year}{2020}\natexlab{f}.
\newblock \bibinfo{title}{Turi Create}.
\newblock
\newblock
\urldef\tempurl%
\url{https://github.com/apple/turicreate/}
\showURL{%
\tempurl}


\bibitem[\protect\citeauthoryear{Apple}{Apple}{2020g}]%
        {uiaccessibility}
\bibfield{author}{\bibinfo{person}{Apple}.} \bibinfo{year}{2020}\natexlab{g}.
\newblock \bibinfo{title}{UIAccessibility}.
\newblock
\newblock
\urldef\tempurl%
\url{https://developer.apple.com/documentation/objectivec/vision/nsobject/uiaccessibility/}
\showURL{%
\tempurl}


\bibitem[\protect\citeauthoryear{Apple}{Apple}{2020h}]%
        {switchcontrol}
\bibfield{author}{\bibinfo{person}{Apple}.} \bibinfo{year}{2020}\natexlab{h}.
\newblock \bibinfo{title}{Use Switch Control to navigate your iPhone, iPad, or
  iPod touch}.
\newblock
\newblock
\newblock
\shownote{https://support.apple.com/en-us/HT201370.}


\bibitem[\protect\citeauthoryear{Apple}{Apple}{2020i}]%
        {voicecontrol}
\bibfield{author}{\bibinfo{person}{Apple}.} \bibinfo{year}{2020}\natexlab{i}.
\newblock \bibinfo{title}{Use Voice Control on your iPhone, iPad, or iPod
  touch}.
\newblock
\newblock
\newblock
\shownote{https://support.apple.com/en-us/HT210417.}


\bibitem[\protect\citeauthoryear{Apple}{Apple}{2020j}]%
        {appleVoiceOver}
\bibfield{author}{\bibinfo{person}{Apple}.} \bibinfo{year}{2020}\natexlab{j}.
\newblock \bibinfo{title}{Vision Accessibility - iPhone}.
\newblock
\newblock
\urldef\tempurl%
\url{https://www.apple.com/accessibility/iphone/vision/}
\showURL{%
\tempurl}


\bibitem[\protect\citeauthoryear{Banovic, Grossman, Matejka, and
  Fitzmaurice}{Banovic et~al\mbox{.}}{2012}]%
        {banovicWaken2012}
\bibfield{author}{\bibinfo{person}{Nikola Banovic}, \bibinfo{person}{Tovi
  Grossman}, \bibinfo{person}{Justin Matejka}, {and} \bibinfo{person}{George
  Fitzmaurice}.} \bibinfo{year}{2012}\natexlab{}.
\newblock \showarticletitle{Waken: {Reverse} {Engineering} {Usage}
  {Information} and {Interface} {Structure} from {Software} {Videos}}. In
  \bibinfo{booktitle}{\emph{Proceedings of the 25th {Annual} {ACM} {Symposium}
  on {User} {Interface} {Software} and {Technology}}}
  \emph{(\bibinfo{series}{{UIST} '12})}. \bibinfo{publisher}{ACM},
  \bibinfo{address}{New York, NY, USA}, \bibinfo{pages}{83--92}.
\newblock


\bibitem[\protect\citeauthoryear{Beltramelli}{Beltramelli}{2018}]%
        {beltramelli2017pix2code}
\bibfield{author}{\bibinfo{person}{Tony Beltramelli}.}
  \bibinfo{year}{2018}\natexlab{}.
\newblock \showarticletitle{Pix2Code: Generating Code from a Graphical User
  Interface Screenshot}. In \bibinfo{booktitle}{\emph{Proceedings of the ACM
  SIGCHI Symposium on Engineering Interactive Computing Systems}} (Paris,
  France) \emph{(\bibinfo{series}{EICS '18})}. \bibinfo{publisher}{ACM},
  \bibinfo{address}{New York, NY, USA}, Article \bibinfo{articleno}{3},
  \bibinfo{numpages}{6}~pages.
\newblock
\showISBNx{978-1-4503-5897-2}


\bibitem[\protect\citeauthoryear{Bielik, Fischer, and Vechev}{Bielik
  et~al\mbox{.}}{2018}]%
        {bielik2018robust}
\bibfield{author}{\bibinfo{person}{Pavol Bielik}, \bibinfo{person}{Marc
  Fischer}, {and} \bibinfo{person}{Martin Vechev}.}
  \bibinfo{year}{2018}\natexlab{}.
\newblock \showarticletitle{Robust relational layout synthesis from examples
  for Android}.
\newblock \bibinfo{journal}{\emph{Proceedings of the ACM on Programming
  Languages}} \bibinfo{volume}{2}, \bibinfo{number}{OOPSLA}
  (\bibinfo{year}{2018}), \bibinfo{pages}{1--29}.
\newblock


\bibitem[\protect\citeauthoryear{Brajnik}{Brajnik}{2008}]%
        {brajnik2008comparative}
\bibfield{author}{\bibinfo{person}{Giorgio Brajnik}.}
  \bibinfo{year}{2008}\natexlab{}.
\newblock \showarticletitle{A comparative test of web accessibility evaluation
  methods}. In \bibinfo{booktitle}{\emph{Proceedings of the 10th international
  ACM SIGACCESS conference on Computers and accessibility}}.
  \bibinfo{pages}{113--120}.
\newblock


\bibitem[\protect\citeauthoryear{Caldwell, Cooper, Reid, Vanderheiden,
  Chisholm, Slatin, and White}{Caldwell et~al\mbox{.}}{2008}]%
        {wcag}
\bibfield{author}{\bibinfo{person}{Ben Caldwell}, \bibinfo{person}{Michael
  Cooper}, \bibinfo{person}{Loretta~Guarino Reid}, \bibinfo{person}{Gregg
  Vanderheiden}, \bibinfo{person}{Wendy Chisholm}, \bibinfo{person}{John
  Slatin}, {and} \bibinfo{person}{Jason White}.}
  \bibinfo{year}{2008}\natexlab{}.
\newblock \showarticletitle{Web content accessibility guidelines (WCAG) 2.0}.
\newblock \bibinfo{journal}{\emph{WWW Consortium (W3C)}}
  (\bibinfo{year}{2008}).
\newblock


\bibitem[\protect\citeauthoryear{Carter and Fourney}{Carter and
  Fourney}{2007}]%
        {carter2007techniques}
\bibfield{author}{\bibinfo{person}{Jim~A Carter} {and} \bibinfo{person}{David~W
  Fourney}.} \bibinfo{year}{2007}\natexlab{}.
\newblock \showarticletitle{Techniques to assist in developing accessibility
  engineers}. In \bibinfo{booktitle}{\emph{Proceedings of the 9th International
  ACM SIGACCESS Conference on Computers and accessibility}}.
  \bibinfo{pages}{123--130}.
\newblock


\bibitem[\protect\citeauthoryear{Chang, Yeh, and Miller}{Chang
  et~al\mbox{.}}{2011}]%
        {changeAssociating2011}
\bibfield{author}{\bibinfo{person}{Tsung-Hsiang Chang}, \bibinfo{person}{Tom
  Yeh}, {and} \bibinfo{person}{Rob Miller}.} \bibinfo{year}{2011}\natexlab{}.
\newblock \showarticletitle{Associating the {Visual} {Representation} of {User}
  {Interfaces} {With} {Their} {Internal} {Structures} and {Metadata}}. In
  \bibinfo{booktitle}{\emph{Proceedings of the 24th {Annual} {ACM} {Symposium}
  on {User} {Interface} {Software} and {Technology}}}
  \emph{(\bibinfo{series}{{UIST} '11})}. \bibinfo{publisher}{ACM},
  \bibinfo{address}{New York, New York, USA}.
\newblock


\bibitem[\protect\citeauthoryear{Checker}{Checker}{2011}]%
        {checker2011idi}
\bibfield{author}{\bibinfo{person}{Web~Accessibility Checker}.}
  \bibinfo{year}{2011}\natexlab{}.
\newblock \showarticletitle{IDI Web Accessibility Checker: Web Accessibility
  Checker}.
\newblock \bibinfo{journal}{\emph{Online. Available:
  http://achecker.ca/checker/index. php. [Last accessed: 28 10 2014]}}
  (\bibinfo{year}{2011}).
\newblock


\bibitem[\protect\citeauthoryear{Chen, Feng, Liu, Xing, and Zhao}{Chen
  et~al\mbox{.}}{2020c}]%
        {chen2020tags}
\bibfield{author}{\bibinfo{person}{Chunyang Chen}, \bibinfo{person}{Sidong
  Feng}, \bibinfo{person}{Zhengyang Liu}, \bibinfo{person}{Zhenchang Xing},
  {and} \bibinfo{person}{Shengdong Zhao}.} \bibinfo{year}{2020}\natexlab{c}.
\newblock \showarticletitle{From Lost to Found: Discover Missing UI Design
  Semantics through Recovering Missing Tags}.
\newblock \bibinfo{journal}{\emph{Proceedings of the ACM on Human-Computer
  Interaction}} \bibinfo{number}{CSCW} (\bibinfo{year}{2020}).
\newblock


\bibitem[\protect\citeauthoryear{Chen, Feng, Xing, Liu, Zhao, and Wang}{Chen
  et~al\mbox{.}}{2019c}]%
        {chen2019gallery}
\bibfield{author}{\bibinfo{person}{Chunyang Chen}, \bibinfo{person}{Sidong
  Feng}, \bibinfo{person}{Zhenchang Xing}, \bibinfo{person}{Linda Liu},
  \bibinfo{person}{Shengdong Zhao}, {and} \bibinfo{person}{Jinshui Wang}.}
  \bibinfo{year}{2019}\natexlab{c}.
\newblock \showarticletitle{Gallery {D}.{C}.: {Design} {Search} and {Knowledge}
  {Discovery} {Through} {Auto}-created {GUI} {Component} {Gallery}}.
\newblock \bibinfo{journal}{\emph{Proceedings of the ACM on Human-Computer
  Interaction}} \bibinfo{volume}{3}, \bibinfo{number}{CSCW}
  (\bibinfo{date}{Nov.} \bibinfo{year}{2019}).
\newblock
\showISSN{2573-0142}


\bibitem[\protect\citeauthoryear{Chen, Su, Meng, Xing, and Liu}{Chen
  et~al\mbox{.}}{2018}]%
        {chen2018neural}
\bibfield{author}{\bibinfo{person}{Chunyang Chen}, \bibinfo{person}{Ting Su},
  \bibinfo{person}{Guozhu Meng}, \bibinfo{person}{Zhenchang Xing}, {and}
  \bibinfo{person}{Yang Liu}.} \bibinfo{year}{2018}\natexlab{}.
\newblock \showarticletitle{From UI Design Image to GUI Skeleton: A Neural
  Machine Translator to Bootstrap Mobile GUI Implementation}. In
  \bibinfo{booktitle}{\emph{Proceedings of the 40th {International}
  {Conference} on {Software} {Engineering}}} \emph{(\bibinfo{series}{{ICSE}
  '18})}. \bibinfo{publisher}{ACM}, \bibinfo{address}{New York, NY, USA},
  \bibinfo{pages}{665--676}.
\newblock
\showISBNx{978-1-4503-5638-1}


\bibitem[\protect\citeauthoryear{Chen, Chen, Xing, Xia, Zhu, Grundy, and
  Wang}{Chen et~al\mbox{.}}{2020a}]%
        {chen2020wireframesearch}
\bibfield{author}{\bibinfo{person}{Jieshan Chen}, \bibinfo{person}{Chunyang
  Chen}, \bibinfo{person}{Zhenchang Xing}, \bibinfo{person}{Xin Xia},
  \bibinfo{person}{Liming Zhu}, \bibinfo{person}{John Grundy}, {and}
  \bibinfo{person}{Jinshui Wang}.} \bibinfo{year}{2020}\natexlab{a}.
\newblock \showarticletitle{Wireframe-Based UI Design Search through Image
  Autoencoder}.
\newblock \bibinfo{journal}{\emph{ACM Trans. Softw. Eng. Methodol.}}
  \bibinfo{volume}{29}, \bibinfo{number}{3}, Article \bibinfo{articleno}{19}
  (\bibinfo{date}{June} \bibinfo{year}{2020}), \bibinfo{numpages}{31}~pages.
\newblock
\showISSN{1049-331X}
\urldef\tempurl%
\url{https://doi.org/10.1145/3391613}
\showDOI{\tempurl}


\bibitem[\protect\citeauthoryear{Chen, Chen, Xing, Xu, Zhu, Li, and Wang}{Chen
  et~al\mbox{.}}{2020b}]%
        {chen2020unblind}
\bibfield{author}{\bibinfo{person}{Jieshan Chen}, \bibinfo{person}{Chunyang
  Chen}, \bibinfo{person}{Zhenchang Xing}, \bibinfo{person}{Xiwei Xu},
  \bibinfo{person}{Liming Zhu}, \bibinfo{person}{Guoqiang Li}, {and}
  \bibinfo{person}{Jinshui Wang}.} \bibinfo{year}{2020}\natexlab{b}.
\newblock \showarticletitle{Unblind Your Apps: Predicting Natural-Language
  Labels for Mobile GUI Components by Deep Learning}.
\newblock \bibinfo{journal}{\emph{2020 IEEE/ACM 42nd International Conference
  on Software Engineering (ICSE)}} (\bibinfo{year}{2020}).
\newblock


\bibitem[\protect\citeauthoryear{Chen, Xie, Xing, Chen, Xu, and Zhu}{Chen
  et~al\mbox{.}}{2020d}]%
        {chen2020object}
\bibfield{author}{\bibinfo{person}{Jieshan Chen}, \bibinfo{person}{Mulong Xie},
  \bibinfo{person}{Zhenchang Xing}, \bibinfo{person}{Chunyang Chen},
  \bibinfo{person}{Xiwei Xu}, {and} \bibinfo{person}{Liming Zhu}.}
  \bibinfo{year}{2020}\natexlab{d}.
\newblock \showarticletitle{Object Detection for Graphical User Interface: Old
  Fashioned or Deep Learning or a Combination?}
\newblock \bibinfo{journal}{\emph{2020 IEEE/ACM 42nd International Conference
  on Software Engineering (ICSE)}} (\bibinfo{year}{2020}).
\newblock


\bibitem[\protect\citeauthoryear{Chen, Fan, Chen, Xue, Liu, and Xu}{Chen
  et~al\mbox{.}}{2019a}]%
        {chen2019gui}
\bibfield{author}{\bibinfo{person}{Sen Chen}, \bibinfo{person}{Lingling Fan},
  \bibinfo{person}{Chunyang Chen}, \bibinfo{person}{Minhui Xue},
  \bibinfo{person}{Yang Liu}, {and} \bibinfo{person}{Lihua Xu}.}
  \bibinfo{year}{2019}\natexlab{a}.
\newblock \showarticletitle{GUI-Squatting Attack: Automated Generation of
  Android Phishing Apps}.
\newblock \bibinfo{journal}{\emph{IEEE Transactions on Dependable and Secure
  Computing}} (\bibinfo{year}{2019}).
\newblock


\bibitem[\protect\citeauthoryear{Chen, Fan, Su, Ma, Liu, and Xu}{Chen
  et~al\mbox{.}}{2019b}]%
        {chen2019automated}
\bibfield{author}{\bibinfo{person}{Sen Chen}, \bibinfo{person}{Lingling Fan},
  \bibinfo{person}{Ting Su}, \bibinfo{person}{Lei Ma}, \bibinfo{person}{Yang
  Liu}, {and} \bibinfo{person}{Lihua Xu}.} \bibinfo{year}{2019}\natexlab{b}.
\newblock \showarticletitle{Automated cross-platform GUI code generation for
  mobile apps}. In \bibinfo{booktitle}{\emph{2019 IEEE 1st International
  Workshop on Artificial Intelligence for Mobile (AI4Mobile)}}. IEEE,
  \bibinfo{pages}{13--16}.
\newblock


\bibitem[\protect\citeauthoryear{Deka, Huang, Franzen, Hibschman, Afergan, Li,
  Nichols, and Kumar}{Deka et~al\mbox{.}}{2017}]%
        {deka2017rico}
\bibfield{author}{\bibinfo{person}{Biplab Deka}, \bibinfo{person}{Zifeng
  Huang}, \bibinfo{person}{Chad Franzen}, \bibinfo{person}{Joshua Hibschman},
  \bibinfo{person}{Daniel Afergan}, \bibinfo{person}{Yang Li},
  \bibinfo{person}{Jeffrey Nichols}, {and} \bibinfo{person}{Ranjitha Kumar}.}
  \bibinfo{year}{2017}\natexlab{}.
\newblock \showarticletitle{Rico: A mobile app dataset for building data-driven
  design applications}. In \bibinfo{booktitle}{\emph{Proceedings of the 30th
  Annual ACM Symposium on User Interface Software and Technology}}.
\newblock


\bibitem[\protect\citeauthoryear{Deka, Huang, and Kumar}{Deka
  et~al\mbox{.}}{2016}]%
        {deka2016erica}
\bibfield{author}{\bibinfo{person}{Biplab Deka}, \bibinfo{person}{Zifeng
  Huang}, {and} \bibinfo{person}{Ranjitha Kumar}.}
  \bibinfo{year}{2016}\natexlab{}.
\newblock \showarticletitle{ERICA: Interaction Mining Mobile Apps}. In
  \bibinfo{booktitle}{\emph{Proceedings of the 29th Annual Symposium on User
  Interface Software and Technology}} (Tokyo, Japan)
  \emph{(\bibinfo{series}{UIST '16})}. \bibinfo{publisher}{Association for
  Computing Machinery}, \bibinfo{address}{New York, NY, USA},
  \bibinfo{pages}{767–776}.
\newblock
\showISBNx{9781450341899}
\urldef\tempurl%
\url{https://doi.org/10.1145/2984511.2984581}
\showDOI{\tempurl}


\bibitem[\protect\citeauthoryear{Dell, Vaidyanathan, Medhi, Cutrell, and
  Thies}{Dell et~al\mbox{.}}{2012}]%
        {dell2012yours}
\bibfield{author}{\bibinfo{person}{Nicola Dell}, \bibinfo{person}{Vidya
  Vaidyanathan}, \bibinfo{person}{Indrani Medhi}, \bibinfo{person}{Edward
  Cutrell}, {and} \bibinfo{person}{William Thies}.}
  \bibinfo{year}{2012}\natexlab{}.
\newblock \showarticletitle{"Yours is Better!" Participant Response Bias in
  HCI}. In \bibinfo{booktitle}{\emph{Proceedings of the SIGCHI Conference on
  Human Factors in Computing Systems}}. \bibinfo{pages}{1321--1330}.
\newblock


\bibitem[\protect\citeauthoryear{Dixon and Fogarty}{Dixon and Fogarty}{2010}]%
        {dixon2010prefab}
\bibfield{author}{\bibinfo{person}{Morgan Dixon} {and} \bibinfo{person}{James
  Fogarty}.} \bibinfo{year}{2010}\natexlab{}.
\newblock \showarticletitle{Prefab: Implementing Advanced Behaviors Using
  Pixel-Based Reverse Engineering of Interface Structure}. In
  \bibinfo{booktitle}{\emph{Proceedings of the SIGCHI Conference on Human
  Factors in Computing Systems}} (Atlanta, Georgia, USA)
  \emph{(\bibinfo{series}{CHI ’10})}. \bibinfo{publisher}{Association for
  Computing Machinery}, \bibinfo{address}{New York, NY, USA},
  \bibinfo{pages}{1525–1534}.
\newblock
\showISBNx{9781605589299}
\urldef\tempurl%
\url{https://doi.org/10.1145/1753326.1753554}
\showDOI{\tempurl}


\bibitem[\protect\citeauthoryear{Dixon, Fogarty, and Wobbrock}{Dixon
  et~al\mbox{.}}{2012}]%
        {dixon2012bubblecursor}
\bibfield{author}{\bibinfo{person}{Morgan Dixon}, \bibinfo{person}{James
  Fogarty}, {and} \bibinfo{person}{Jacob Wobbrock}.}
  \bibinfo{year}{2012}\natexlab{}.
\newblock \showarticletitle{A General-Purpose Target-Aware Pointing Enhancement
  Using Pixel-Level Analysis of Graphical Interfaces}. In
  \bibinfo{booktitle}{\emph{Proceedings of the SIGCHI Conference on Human
  Factors in Computing Systems}} (Austin, Texas, USA)
  \emph{(\bibinfo{series}{CHI '12})}. \bibinfo{publisher}{Association for
  Computing Machinery}, \bibinfo{address}{New York, NY, USA},
  \bibinfo{pages}{3167–3176}.
\newblock
\showISBNx{9781450310154}
\urldef\tempurl%
\url{https://doi.org/10.1145/2207676.2208734}
\showDOI{\tempurl}


\bibitem[\protect\citeauthoryear{Eagan, Beaudouin-Lafon, and Mackay}{Eagan
  et~al\mbox{.}}{2011}]%
        {eagan_cracking_2011}
\bibfield{author}{\bibinfo{person}{James~R. Eagan}, \bibinfo{person}{Michel
  Beaudouin-Lafon}, {and} \bibinfo{person}{Wendy~E. Mackay}.}
  \bibinfo{year}{2011}\natexlab{}.
\newblock \showarticletitle{Cracking the {Cocoa} {Nut}: {User} {Interface}
  {Programming} at {Runtime}}. In \bibinfo{booktitle}{\emph{Proceedings of the
  24th {Annual} {ACM} {Symposium} on {User} {Interface} {Software} and
  {Technology}}} \emph{(\bibinfo{series}{{UIST} '11})}.
  \bibinfo{publisher}{ACM}, \bibinfo{address}{New York, NY, USA}.
\newblock
\showISBNx{978-1-4503-0716-1}


\bibitem[\protect\citeauthoryear{Everingham, Van~Gool, Williams, Winn, and
  Zisserman}{Everingham et~al\mbox{.}}{2010}]%
        {everingham2010pascal}
\bibfield{author}{\bibinfo{person}{Mark Everingham}, \bibinfo{person}{Luc
  Van~Gool}, \bibinfo{person}{Christopher~KI Williams}, \bibinfo{person}{John
  Winn}, {and} \bibinfo{person}{Andrew Zisserman}.}
  \bibinfo{year}{2010}\natexlab{}.
\newblock \showarticletitle{The Pascal Visual Object Classes (voc) Challenge}.
\newblock \bibinfo{journal}{\emph{International Journal of Computer Vision}}
  \bibinfo{volume}{88}, \bibinfo{number}{2} (\bibinfo{year}{2010}),
  \bibinfo{pages}{303--338}.
\newblock


\bibitem[\protect\citeauthoryear{Games}{Games}{2020}]%
        {unityPlugin}
\bibfield{author}{\bibinfo{person}{MetalPop Games}.}
  \bibinfo{year}{2020}\natexlab{}.
\newblock \bibinfo{title}{UI Accessibility Plugin (UAP): GUI Tools: Unity Asset
  Store}.
\newblock
\newblock
\urldef\tempurl%
\url{https://assetstore.unity.com/packages/tools/gui/ui-accessibility-plugin-uap-87935}
\showURL{%
\tempurl}


\bibitem[\protect\citeauthoryear{Gleason, Pavel, McCamey, Low, Carrington,
  Kitani, and Bigham}{Gleason et~al\mbox{.}}{2020}]%
        {gleason2020twitter}
\bibfield{author}{\bibinfo{person}{Cole Gleason}, \bibinfo{person}{Amy Pavel},
  \bibinfo{person}{Emma McCamey}, \bibinfo{person}{Christina Low},
  \bibinfo{person}{Patrick Carrington}, \bibinfo{person}{Kris~M Kitani}, {and}
  \bibinfo{person}{Jeffrey~P Bigham}.} \bibinfo{year}{2020}\natexlab{}.
\newblock \showarticletitle{Twitter A11y: A Browser Extension to Make Twitter
  Images Accessible}. In \bibinfo{booktitle}{\emph{Proceedings of the 2020 CHI
  Conference on Human Factors in Computing Systems}}. \bibinfo{pages}{1--12}.
\newblock


\bibitem[\protect\citeauthoryear{Google}{Google}{2020a}]%
        {switchaccess}
\bibfield{author}{\bibinfo{person}{Google}.} \bibinfo{year}{2020}\natexlab{a}.
\newblock \bibinfo{title}{Android Accessibility Help: Switch Access}.
\newblock
\newblock
\newblock
\shownote{https://support.google.com/accessibility/android/topic/6151780.}


\bibitem[\protect\citeauthoryear{Google}{Google}{2020b}]%
        {androidax}
\bibfield{author}{\bibinfo{person}{Google}.} \bibinfo{year}{2020}\natexlab{b}.
\newblock \bibinfo{title}{Build more accessible apps}.
\newblock
\newblock
\newblock
\shownote{https://developer.android.com/guide/topics/ui/accessibility/.}


\bibitem[\protect\citeauthoryear{Google}{Google}{2020c}]%
        {googleTalkback}
\bibfield{author}{\bibinfo{person}{Google}.} \bibinfo{year}{2020}\natexlab{c}.
\newblock \bibinfo{title}{Get started on Android with TalkBack - Android
  Accessibility Help}.
\newblock
\newblock
\urldef\tempurl%
\url{https://support.google.com/accessibility/android/answer/6283677?hl=en}
\showURL{%
\tempurl}


\bibitem[\protect\citeauthoryear{Guinness, Cutrell, and Morris}{Guinness
  et~al\mbox{.}}{2018}]%
        {guinness2018caption}
\bibfield{author}{\bibinfo{person}{Darren Guinness}, \bibinfo{person}{Edward
  Cutrell}, {and} \bibinfo{person}{Meredith~Ringel Morris}.}
  \bibinfo{year}{2018}\natexlab{}.
\newblock \showarticletitle{Caption crawler: Enabling reusable alternative text
  descriptions using reverse image search}. In
  \bibinfo{booktitle}{\emph{Proceedings of the 2018 CHI Conference on Human
  Factors in Computing Systems}}. \bibinfo{pages}{1--11}.
\newblock


\bibitem[\protect\citeauthoryear{Gurari, Zhao, Zhang, and Bhattacharya}{Gurari
  et~al\mbox{.}}{2020}]%
        {gurari2020captioning}
\bibfield{author}{\bibinfo{person}{Danna Gurari}, \bibinfo{person}{Yinan Zhao},
  \bibinfo{person}{Meng Zhang}, {and} \bibinfo{person}{Nilavra Bhattacharya}.}
  \bibinfo{year}{2020}\natexlab{}.
\newblock \showarticletitle{Captioning Images Taken by People Who Are Blind}.
\newblock \bibinfo{journal}{\emph{arXiv preprint arXiv:2002.08565}}
  (\bibinfo{year}{2020}).
\newblock


\bibitem[\protect\citeauthoryear{Ha, Haralick, and Phillips}{Ha
  et~al\mbox{.}}{1995}]%
        {ha1995recursive}
\bibfield{author}{\bibinfo{person}{Jaekyu Ha}, \bibinfo{person}{Robert~M
  Haralick}, {and} \bibinfo{person}{Ihsin~T Phillips}.}
  \bibinfo{year}{1995}\natexlab{}.
\newblock \showarticletitle{Recursive XY Cut Using Bounding Boxes of Connected
  Components}. In \bibinfo{booktitle}{\emph{Proceedings of 3rd International
  Conference on Document Analysis and Recognition}}
  \emph{(\bibinfo{series}{ICDAR '95}, Vol.~\bibinfo{volume}{2})}. IEEE,
  \bibinfo{pages}{952--955}.
\newblock


\bibitem[\protect\citeauthoryear{Hanson and Richards}{Hanson and
  Richards}{2013}]%
        {hanson2013progress}
\bibfield{author}{\bibinfo{person}{Vicki~L Hanson} {and}
  \bibinfo{person}{John~T Richards}.} \bibinfo{year}{2013}\natexlab{}.
\newblock \showarticletitle{Progress on website accessibility?}
\newblock \bibinfo{journal}{\emph{ACM Transactions on the Web (TWEB)}}
  \bibinfo{volume}{7}, \bibinfo{number}{1} (\bibinfo{year}{2013}),
  \bibinfo{pages}{1--30}.
\newblock


\bibitem[\protect\citeauthoryear{He, Gkioxari, Dollar, and Girshick}{He
  et~al\mbox{.}}{2017}]%
        {he2017mask}
\bibfield{author}{\bibinfo{person}{K He}, \bibinfo{person}{G Gkioxari},
  \bibinfo{person}{P Dollar}, {and} \bibinfo{person}{R Girshick}.}
  \bibinfo{year}{2017}\natexlab{}.
\newblock \showarticletitle{Mask R-CNN}. In \bibinfo{booktitle}{\emph{2017 IEEE
  International Conference on Computer Vision (ICCV)}}.
  \bibinfo{pages}{2980--2988}.
\newblock


\bibitem[\protect\citeauthoryear{Hossain, Sohel, Shiratuddin, and Laga}{Hossain
  et~al\mbox{.}}{2019}]%
        {hossain2019comprehensive}
\bibfield{author}{\bibinfo{person}{MD~Zakir Hossain}, \bibinfo{person}{Ferdous
  Sohel}, \bibinfo{person}{Mohd~Fairuz Shiratuddin}, {and}
  \bibinfo{person}{Hamid Laga}.} \bibinfo{year}{2019}\natexlab{}.
\newblock \showarticletitle{A comprehensive survey of deep learning for image
  captioning}.
\newblock \bibinfo{journal}{\emph{ACM Computing Surveys (CSUR)}}
  \bibinfo{volume}{51}, \bibinfo{number}{6} (\bibinfo{year}{2019}),
  \bibinfo{pages}{1--36}.
\newblock


\bibitem[\protect\citeauthoryear{Huang, Canny, and Nichols}{Huang
  et~al\mbox{.}}{2019}]%
        {huang2019swire}
\bibfield{author}{\bibinfo{person}{Forrest Huang}, \bibinfo{person}{John~F
  Canny}, {and} \bibinfo{person}{Jeffrey Nichols}.}
  \bibinfo{year}{2019}\natexlab{}.
\newblock \showarticletitle{Swire: Sketch-based user interface retrieval}. In
  \bibinfo{booktitle}{\emph{Proceedings of the 2019 CHI Conference on Human
  Factors in Computing Systems}}. \bibinfo{pages}{1--10}.
\newblock


\bibitem[\protect\citeauthoryear{Karpathy and Fei-Fei}{Karpathy and
  Fei-Fei}{2015}]%
        {karpathy2015deep}
\bibfield{author}{\bibinfo{person}{Andrej Karpathy} {and} \bibinfo{person}{Li
  Fei-Fei}.} \bibinfo{year}{2015}\natexlab{}.
\newblock \showarticletitle{Deep visual-semantic alignments for generating
  image descriptions}. In \bibinfo{booktitle}{\emph{Proceedings of the IEEE
  conference on computer vision and pattern recognition}}.
  \bibinfo{pages}{3128--3137}.
\newblock


\bibitem[\protect\citeauthoryear{Kaufman, Rosset, Perlich, and
  Stitelman}{Kaufman et~al\mbox{.}}{2012}]%
        {kaufman2012leakage}
\bibfield{author}{\bibinfo{person}{Shachar Kaufman}, \bibinfo{person}{Saharon
  Rosset}, \bibinfo{person}{Claudia Perlich}, {and} \bibinfo{person}{Ori
  Stitelman}.} \bibinfo{year}{2012}\natexlab{}.
\newblock \showarticletitle{Leakage in data mining: Formulation, detection, and
  avoidance}.
\newblock \bibinfo{journal}{\emph{ACM Transactions on Knowledge Discovery from
  Data (TKDD)}} \bibinfo{volume}{6}, \bibinfo{number}{4}
  (\bibinfo{year}{2012}), \bibinfo{pages}{1--21}.
\newblock


\bibitem[\protect\citeauthoryear{Kumar, Talton, Ahmad, and Klemmer}{Kumar
  et~al\mbox{.}}{2011}]%
        {kumar2011bricolage}
\bibfield{author}{\bibinfo{person}{Ranjitha Kumar}, \bibinfo{person}{Jerry~O
  Talton}, \bibinfo{person}{Salman Ahmad}, {and} \bibinfo{person}{Scott~R
  Klemmer}.} \bibinfo{year}{2011}\natexlab{}.
\newblock \showarticletitle{Bricolage: example-based retargeting for web
  design}. In \bibinfo{booktitle}{\emph{Proceedings of the SIGCHI Conference on
  Human Factors in Computing Systems}}. \bibinfo{pages}{2197--2206}.
\newblock


\bibitem[\protect\citeauthoryear{Li, Azaria, and Myers}{Li
  et~al\mbox{.}}{2017}]%
        {sugalite}
\bibfield{author}{\bibinfo{person}{Toby Jia-Jun Li}, \bibinfo{person}{Amos
  Azaria}, {and} \bibinfo{person}{Brad~A. Myers}.}
  \bibinfo{year}{2017}\natexlab{}.
\newblock \showarticletitle{SUGILITE: Creating Multimodal Smartphone Automation
  by Demonstration}. In \bibinfo{booktitle}{\emph{Proceedings of the 2017 CHI
  Conference on Human Factors in Computing Systems}} (Denver, Colorado, USA)
  \emph{(\bibinfo{series}{CHI '17})}. \bibinfo{publisher}{Association for
  Computing Machinery}, \bibinfo{address}{New York, NY, USA},
  \bibinfo{pages}{6038–6049}.
\newblock
\showISBNx{9781450346559}
\urldef\tempurl%
\url{https://doi.org/10.1145/3025453.3025483}
\showDOI{\tempurl}


\bibitem[\protect\citeauthoryear{Liu, Craft, Situ, Yumer, Mech, and Kumar}{Liu
  et~al\mbox{.}}{2018}]%
        {liu2018semantics}
\bibfield{author}{\bibinfo{person}{Thomas~F. Liu}, \bibinfo{person}{Mark
  Craft}, \bibinfo{person}{Jason Situ}, \bibinfo{person}{Ersin Yumer},
  \bibinfo{person}{Radomir Mech}, {and} \bibinfo{person}{Ranjitha Kumar}.}
  \bibinfo{year}{2018}\natexlab{}.
\newblock \showarticletitle{Learning Design Semantics for Mobile Apps}. In
  \bibinfo{booktitle}{\emph{Proceedings of the 31st Annual ACM Symposium on
  User Interface Software and Technology}} (Berlin, Germany)
  \emph{(\bibinfo{series}{UIST '18})}. \bibinfo{publisher}{Association for
  Computing Machinery}, \bibinfo{address}{New York, NY, USA},
  \bibinfo{pages}{569–579}.
\newblock
\showISBNx{9781450359481}
\urldef\tempurl%
\url{https://doi.org/10.1145/3242587.3242650}
\showDOI{\tempurl}


\bibitem[\protect\citeauthoryear{Liu, Anguelov, Erhan, Szegedy, Reed, Fu, and
  Berg}{Liu et~al\mbox{.}}{2016}]%
        {liu2016ssd}
\bibfield{author}{\bibinfo{person}{Wei Liu}, \bibinfo{person}{Dragomir
  Anguelov}, \bibinfo{person}{Dumitru Erhan}, \bibinfo{person}{Christian
  Szegedy}, \bibinfo{person}{Scott Reed}, \bibinfo{person}{Cheng-Yang Fu},
  {and} \bibinfo{person}{Alexander~C Berg}.} \bibinfo{year}{2016}\natexlab{}.
\newblock \showarticletitle{SSD: Single shot multibox detector}. In
  \bibinfo{booktitle}{\emph{European conference on computer vision}}. Springer,
  \bibinfo{pages}{21--37}.
\newblock


\bibitem[\protect\citeauthoryear{Memon, Banerjee, and Nagarajan}{Memon
  et~al\mbox{.}}{2003}]%
        {memon2003gui}
\bibfield{author}{\bibinfo{person}{Atif Memon}, \bibinfo{person}{Ishan
  Banerjee}, {and} \bibinfo{person}{Adithya Nagarajan}.}
  \bibinfo{year}{2003}\natexlab{}.
\newblock \showarticletitle{GUI ripping: Reverse engineering of graphical user
  interfaces for testing}. In \bibinfo{booktitle}{\emph{Proceedings of the 10th
  Working Conference on Reverse Engineering}} \emph{(\bibinfo{series}{WCRE
  '03})}. Citeseer, \bibinfo{pages}{260--269}.
\newblock


\bibitem[\protect\citeauthoryear{Moran, Bernal-C{\'a}rdenas, Curcio, Bonett,
  and Poshyvanyk}{Moran et~al\mbox{.}}{2018}]%
        {moran2018machine}
\bibfield{author}{\bibinfo{person}{Kevin~Patrick Moran},
  \bibinfo{person}{Carlos Bernal-C{\'a}rdenas}, \bibinfo{person}{Michael
  Curcio}, \bibinfo{person}{Richard Bonett}, {and} \bibinfo{person}{Denys
  Poshyvanyk}.} \bibinfo{year}{2018}\natexlab{}.
\newblock \showarticletitle{{Machine Learning-Based Prototyping of Graphical
  User Interfaces for Mobile Apps}}.
\newblock \bibinfo{journal}{\emph{IEEE Transactions on Software Engineering}}
  (\bibinfo{year}{2018}).
\newblock


\bibitem[\protect\citeauthoryear{Nagy and Seth}{Nagy and Seth}{1984}]%
        {nagy1984hierarchical}
\bibfield{author}{\bibinfo{person}{George Nagy} {and}
  \bibinfo{person}{Sharad~C. Seth}.} \bibinfo{year}{1984}\natexlab{}.
\newblock \showarticletitle{Hierarchical Representation of Optically Scanned
  Documents}.
\newblock \bibinfo{journal}{\emph{Proceedings of the 7th International
  Conference on Pattern Recognition}} (\bibinfo{year}{1984}),
  \bibinfo{pages}{347--349}.
\newblock
\urldef\tempurl%
\url{https://ci.nii.ac.jp/naid/10006753131/en/}
\showURL{%
\tempurl}


\bibitem[\protect\citeauthoryear{Neubeck and Van~Gool}{Neubeck and
  Van~Gool}{2006}]%
        {neubeck2006efficient}
\bibfield{author}{\bibinfo{person}{Alexander Neubeck} {and}
  \bibinfo{person}{Luc Van~Gool}.} \bibinfo{year}{2006}\natexlab{}.
\newblock \showarticletitle{Efficient non-maximum suppression}. In
  \bibinfo{booktitle}{\emph{18th International Conference on Pattern
  Recognition (ICPR'06)}}, Vol.~\bibinfo{volume}{3}. IEEE,
  \bibinfo{pages}{850--855}.
\newblock


\bibitem[\protect\citeauthoryear{{Nguyen}, {Vu}, {Pham}, and {Nguyen}}{{Nguyen}
  et~al\mbox{.}}{2018}]%
        {nguyen2018patterns}
\bibfield{author}{\bibinfo{person}{T. {Nguyen}}, \bibinfo{person}{P. {Vu}},
  \bibinfo{person}{H. {Pham}}, {and} \bibinfo{person}{T. {Nguyen}}.}
  \bibinfo{year}{2018}\natexlab{}.
\newblock \showarticletitle{Deep Learning UI Design Patterns of Mobile Apps}.
  In \bibinfo{booktitle}{\emph{2018 IEEE/ACM 40th International Conference on
  Software Engineering: New Ideas and Emerging Technologies Results
  (ICSE-NIER)}}. \bibinfo{pages}{65--68}.
\newblock


\bibitem[\protect\citeauthoryear{Nguyen and Csallner}{Nguyen and
  Csallner}{2015}]%
        {nguyen2015remaui}
\bibfield{author}{\bibinfo{person}{Tuan~Anh Nguyen} {and}
  \bibinfo{person}{Christoph Csallner}.} \bibinfo{year}{2015}\natexlab{}.
\newblock \showarticletitle{Reverse {Engineering} {Mobile} {Application} {User}
  {Interfaces} with {REMAUI} ({T})}. In \bibinfo{booktitle}{\emph{{IEEE}/{ACM}
  {International} {Conference} on {Automated} {Software} {Engineering}}}
  \emph{(\bibinfo{series}{{ASE} '15})}. \bibinfo{publisher}{IEEE},
  \bibinfo{pages}{248--259}.
\newblock


\bibitem[\protect\citeauthoryear{Olsen, Hudson, Verratti, Heiner, and
  Phelps}{Olsen et~al\mbox{.}}{1999}]%
        {olsen_implementing_1999}
\bibfield{author}{\bibinfo{person}{Dan~R. Olsen}, \bibinfo{person}{Scott~E.
  Hudson}, \bibinfo{person}{Thom Verratti}, \bibinfo{person}{Jeremy~M. Heiner},
  {and} \bibinfo{person}{Matt Phelps}.} \bibinfo{year}{1999}\natexlab{}.
\newblock \showarticletitle{Implementing {Interface} {Attachments} based on
  {Surface} {Representations}}. In \bibinfo{booktitle}{\emph{Proceedings of the
  {SIGCHI} {Conference} on {Human} {Factors} in {Computing} {Systems}}}
  \emph{(\bibinfo{series}{{CHI} '99})}. \bibinfo{publisher}{ACM},
  \bibinfo{address}{New York, New York, USA}.
\newblock
\showISBNx{0-201-48559-1}


\bibitem[\protect\citeauthoryear{Pearson, Bailey, and Green}{Pearson
  et~al\mbox{.}}{2011}]%
        {pearson2011tool}
\bibfield{author}{\bibinfo{person}{Elaine Pearson}, \bibinfo{person}{Chrstopher
  Bailey}, {and} \bibinfo{person}{Steve Green}.}
  \bibinfo{year}{2011}\natexlab{}.
\newblock \showarticletitle{A tool to support the web accessibility evaluation
  process for novices}. In \bibinfo{booktitle}{\emph{Proceedings of the 16th
  annual joint conference on Innovation and technology in computer science
  education}}. \bibinfo{pages}{28--32}.
\newblock


\bibitem[\protect\citeauthoryear{Pixelogik}{Pixelogik}{2020}]%
        {colorCube}
\bibfield{author}{\bibinfo{person}{Pixelogik}.}
  \bibinfo{year}{2020}\natexlab{}.
\newblock \bibinfo{title}{ColorCube}.
\newblock
\newblock
\urldef\tempurl%
\url{https://github.com/pixelogik/ColorCube}
\showURL{%
\tempurl}


\bibitem[\protect\citeauthoryear{Pongnumkul, Dontcheva, Li, Wang, Bourdev,
  Avidan, and Cohen}{Pongnumkul et~al\mbox{.}}{2011}]%
        {pongnumkul2011pause}
\bibfield{author}{\bibinfo{person}{Suporn Pongnumkul}, \bibinfo{person}{Mira
  Dontcheva}, \bibinfo{person}{Wilmot Li}, \bibinfo{person}{Jue Wang},
  \bibinfo{person}{Lubomir Bourdev}, \bibinfo{person}{Shai Avidan}, {and}
  \bibinfo{person}{Michael~F Cohen}.} \bibinfo{year}{2011}\natexlab{}.
\newblock \showarticletitle{Pause-and-play: automatically linking screencast
  video tutorials with applications}. In \bibinfo{booktitle}{\emph{Proceedings
  of the 24th annual ACM symposium on User interface software and technology}}.
  \bibinfo{pages}{135--144}.
\newblock


\bibitem[\protect\citeauthoryear{Power, Freire, Petrie, and Swallow}{Power
  et~al\mbox{.}}{2012}]%
        {power2012guidelines}
\bibfield{author}{\bibinfo{person}{Christopher Power},
  \bibinfo{person}{Andr{\'e} Freire}, \bibinfo{person}{Helen Petrie}, {and}
  \bibinfo{person}{David Swallow}.} \bibinfo{year}{2012}\natexlab{}.
\newblock \showarticletitle{Guidelines are only half of the story:
  accessibility problems encountered by blind users on the web}. In
  \bibinfo{booktitle}{\emph{Proceedings of the SIGCHI conference on human
  factors in computing systems}}. \bibinfo{pages}{433--442}.
\newblock


\bibitem[\protect\citeauthoryear{Redmon, Divvala, Girshick, and Farhadi}{Redmon
  et~al\mbox{.}}{2016}]%
        {redmon2016you}
\bibfield{author}{\bibinfo{person}{Joseph Redmon}, \bibinfo{person}{Santosh
  Divvala}, \bibinfo{person}{Ross Girshick}, {and} \bibinfo{person}{Ali
  Farhadi}.} \bibinfo{year}{2016}\natexlab{}.
\newblock \showarticletitle{You only look once: Unified, real-time object
  detection}. In \bibinfo{booktitle}{\emph{Proceedings of the IEEE conference
  on computer vision and pattern recognition}}. \bibinfo{pages}{779--788}.
\newblock


\bibitem[\protect\citeauthoryear{Ren, He, Girshick, and Sun}{Ren
  et~al\mbox{.}}{2015}]%
        {ren2015faster}
\bibfield{author}{\bibinfo{person}{Shaoqing Ren}, \bibinfo{person}{Kaiming He},
  \bibinfo{person}{Ross Girshick}, {and} \bibinfo{person}{Jian Sun}.}
  \bibinfo{year}{2015}\natexlab{}.
\newblock \showarticletitle{Faster r-cnn: Towards real-time object detection
  with region proposal networks}. In \bibinfo{booktitle}{\emph{Advances in
  neural information processing systems}}. \bibinfo{pages}{91--99}.
\newblock


\bibitem[\protect\citeauthoryear{Ross, Zhang, Fogarty, and Wobbrock}{Ross
  et~al\mbox{.}}{2017}]%
        {ross2017epidemiology}
\bibfield{author}{\bibinfo{person}{Anne~Spencer Ross}, \bibinfo{person}{Xiaoyi
  Zhang}, \bibinfo{person}{James Fogarty}, {and} \bibinfo{person}{Jacob~O.
  Wobbrock}.} \bibinfo{year}{2017}\natexlab{}.
\newblock \showarticletitle{Epidemiology as a Framework for Large-Scale Mobile
  Application Accessibility Assessment}. In
  \bibinfo{booktitle}{\emph{Proceedings of the 19th International ACM SIGACCESS
  Conference on Computers and Accessibility}} (Baltimore, Maryland, USA)
  \emph{(\bibinfo{series}{ASSETS ’17})}. \bibinfo{publisher}{Association for
  Computing Machinery}, \bibinfo{address}{New York, NY, USA},
  \bibinfo{pages}{2–11}.
\newblock
\showISBNx{9781450349260}
\urldef\tempurl%
\url{https://doi.org/10.1145/3132525.3132547}
\showDOI{\tempurl}


\bibitem[\protect\citeauthoryear{Ross, Zhang, Fogarty, and Wobbrock}{Ross
  et~al\mbox{.}}{2018}]%
        {ross2018epidemiology}
\bibfield{author}{\bibinfo{person}{Anne~Spencer Ross}, \bibinfo{person}{Xiaoyi
  Zhang}, \bibinfo{person}{James Fogarty}, {and} \bibinfo{person}{Jacob~O.
  Wobbrock}.} \bibinfo{year}{2018}\natexlab{}.
\newblock \showarticletitle{Examining Image-Based Button Labeling for
  Accessibility in Android Apps through Large-Scale Analysis}. In
  \bibinfo{booktitle}{\emph{Proceedings of the 20th International ACM SIGACCESS
  Conference on Computers and Accessibility}} (Galway, Ireland)
  \emph{(\bibinfo{series}{ASSETS ’18})}. \bibinfo{publisher}{Association for
  Computing Machinery}, \bibinfo{address}{New York, NY, USA},
  \bibinfo{pages}{119–130}.
\newblock
\showISBNx{9781450356503}
\urldef\tempurl%
\url{https://doi.org/10.1145/3234695.3236364}
\showDOI{\tempurl}


\bibitem[\protect\citeauthoryear{Ross, Zhang, Fogarty, and Wobbrock}{Ross
  et~al\mbox{.}}{2020}]%
        {ross2020epidemiology}
\bibfield{author}{\bibinfo{person}{Anne~Spencer Ross}, \bibinfo{person}{Xiaoyi
  Zhang}, \bibinfo{person}{James Fogarty}, {and} \bibinfo{person}{Jacob~O.
  Wobbrock}.} \bibinfo{year}{2020}\natexlab{}.
\newblock \showarticletitle{An Epidemiology-Inspired Large-Scale Analysis of
  Android App Accessibility}.
\newblock \bibinfo{journal}{\emph{ACM Trans. Access. Comput.}}
  \bibinfo{volume}{13}, \bibinfo{number}{1}, Article \bibinfo{articleno}{4}
  (\bibinfo{date}{April} \bibinfo{year}{2020}), \bibinfo{numpages}{36}~pages.
\newblock
\showISSN{1936-7228}
\urldef\tempurl%
\url{https://doi.org/10.1145/3348797}
\showDOI{\tempurl}


\bibitem[\protect\citeauthoryear{Santiago}{Santiago}{2018}]%
        {santiago2018}
\bibfield{author}{\bibinfo{person}{Santiago}.} \bibinfo{year}{2018}\natexlab{}.
\newblock \bibinfo{title}{Confusion Matrix in Object Detection with
  TensorFlow}.
\newblock
\newblock
\urldef\tempurl%
\url{https://towardsdatascience.com/confusion-matrix-in-object-detection-with-tensorflow-b9640a927285}
\showURL{%
\tempurl}


\bibitem[\protect\citeauthoryear{Schwerdtfeger}{Schwerdtfeger}{1991}]%
        {outspoken}
\bibfield{author}{\bibinfo{person}{Richard~S. Schwerdtfeger}.}
  \bibinfo{year}{1991}\natexlab{}.
\newblock \bibinfo{title}{Making the GUI Talk}.
\newblock
\newblock
\newblock
\shownote{ftp://service.boulder.ibm.com/sns/sr-os2/sr2doc/guitalk.txt.}


\bibitem[\protect\citeauthoryear{Swearngin, Dontcheva, Li, Brandt, Dixon, and
  Ko}{Swearngin et~al\mbox{.}}{2018}]%
        {swearngin2018rewire}
\bibfield{author}{\bibinfo{person}{Amanda Swearngin}, \bibinfo{person}{Mira
  Dontcheva}, \bibinfo{person}{Wilmot Li}, \bibinfo{person}{Joel Brandt},
  \bibinfo{person}{Morgan Dixon}, {and} \bibinfo{person}{Amy~J. Ko}.}
  \bibinfo{year}{2018}\natexlab{}.
\newblock \showarticletitle{Rewire: {Interface} {Design} {Assistance} from
  {Examples}}. In \bibinfo{booktitle}{\emph{Proceedings of the SIGCHI
  {Conference} on {Human} {Factors} in {Computing} {Systems}}}
  \emph{(\bibinfo{series}{{CHI} '18})}. \bibinfo{publisher}{ACM},
  \bibinfo{address}{New York, NY, USA}, \bibinfo{pages}{504:1--504:12}.
\newblock
\showISBNx{978-1-4503-5620-6}


\bibitem[\protect\citeauthoryear{Swearngin and Li}{Swearngin and Li}{2019}]%
        {swearngin2018tappability}
\bibfield{author}{\bibinfo{person}{Amanda Swearngin} {and}
  \bibinfo{person}{Yang Li}.} \bibinfo{year}{2019}\natexlab{}.
\newblock \showarticletitle{Modeling Mobile Interface Tappability Using
  Crowdsourcing and Deep Learning}. In \bibinfo{booktitle}{\emph{Proceedings of
  the 2019 CHI Conference on Human Factors in Computing Systems}} (Glasgow,
  Scotland Uk) \emph{(\bibinfo{series}{CHI '19})}.
  \bibinfo{publisher}{Association for Computing Machinery},
  \bibinfo{address}{New York, NY, USA}, \bibinfo{pages}{1–11}.
\newblock
\showISBNx{9781450359702}
\urldef\tempurl%
\url{https://doi.org/10.1145/3290605.3300305}
\showDOI{\tempurl}


\bibitem[\protect\citeauthoryear{Takagi, Kawanaka, Kobayashi, Itoh, and
  Asakawa}{Takagi et~al\mbox{.}}{2008}]%
        {takagi2008social}
\bibfield{author}{\bibinfo{person}{Hironobu Takagi}, \bibinfo{person}{Shinya
  Kawanaka}, \bibinfo{person}{Masatomo Kobayashi}, \bibinfo{person}{Takashi
  Itoh}, {and} \bibinfo{person}{Chieko Asakawa}.}
  \bibinfo{year}{2008}\natexlab{}.
\newblock \showarticletitle{Social accessibility: achieving accessibility
  through collaborative metadata authoring}. In
  \bibinfo{booktitle}{\emph{Proceedings of the 10th international ACM SIGACCESS
  conference on Computers and accessibility}}.
\newblock


\bibitem[\protect\citeauthoryear{Tech}{Tech}{2019}]%
        {tech2019ui2code}
\bibfield{author}{\bibinfo{person}{Alibaba Tech}.}
  \bibinfo{year}{2019}\natexlab{}.
\newblock \bibinfo{title}{{UI2code}: {How} to {Fine}-tune {Background} and
  {Foreground} {Analysis}}.
\newblock
\newblock
\urldef\tempurl%
\url{https://medium.com/hackernoon/ui2code-how-to-fine-tune-background-and-foreground-analysis-fb269edcd12c}
\showURL{%
\tempurl}


\bibitem[\protect\citeauthoryear{Trewin, Marques, and Guerreiro}{Trewin
  et~al\mbox{.}}{2015}]%
        {likertbias}
\bibfield{author}{\bibinfo{person}{Shari Trewin}, \bibinfo{person}{Diogo
  Marques}, {and} \bibinfo{person}{Tiago Guerreiro}.}
  \bibinfo{year}{2015}\natexlab{}.
\newblock \showarticletitle{Usage of Subjective Scales in Accessibility
  Research} \emph{(\bibinfo{series}{ASSETS '15})}.
  \bibinfo{publisher}{Association for Computing Machinery},
  \bibinfo{address}{New York, NY, USA}, \bibinfo{pages}{59–67}.
\newblock
\showISBNx{9781450334006}
\urldef\tempurl%
\url{https://doi.org/10.1145/2700648.2809867}
\showDOI{\tempurl}


\bibitem[\protect\citeauthoryear{White, Fraser, and Brown}{White
  et~al\mbox{.}}{2019}]%
        {white2019improving}
\bibfield{author}{\bibinfo{person}{Thomas~D White}, \bibinfo{person}{Gordon
  Fraser}, {and} \bibinfo{person}{Guy~J Brown}.}
  \bibinfo{year}{2019}\natexlab{}.
\newblock \showarticletitle{Improving random GUI testing with image-based
  widget detection}. In \bibinfo{booktitle}{\emph{Proceedings of the 28th ACM
  SIGSOFT International Symposium on Software Testing and Analysis}}.
  \bibinfo{pages}{307--317}.
\newblock


\bibitem[\protect\citeauthoryear{Xi, Yang, Xiao, Yao, Xiong, Xu, Wang, Gao,
  Liu, Xu, and Lu}{Xi et~al\mbox{.}}{2019}]%
        {xi2019deepintent}
\bibfield{author}{\bibinfo{person}{Shengqu Xi}, \bibinfo{person}{Shao Yang},
  \bibinfo{person}{Xusheng Xiao}, \bibinfo{person}{Yuan Yao},
  \bibinfo{person}{Yayuan Xiong}, \bibinfo{person}{Fengyuan Xu},
  \bibinfo{person}{Haoyu Wang}, \bibinfo{person}{Peng Gao},
  \bibinfo{person}{Zhuotao Liu}, \bibinfo{person}{Feng Xu}, {and}
  \bibinfo{person}{Jian Lu}.} \bibinfo{year}{2019}\natexlab{}.
\newblock \showarticletitle{DeepIntent: Deep Icon-Behavior Learning for
  Detecting Intention-Behavior Discrepancy in Mobile Apps}. In
  \bibinfo{booktitle}{\emph{Proceedings of the 2019 ACM SIGSAC Conference on
  Computer and Communications Security}} (London, United Kingdom)
  \emph{(\bibinfo{series}{CCS '19})}. \bibinfo{publisher}{Association for
  Computing Machinery}, \bibinfo{address}{New York, NY, USA},
  \bibinfo{pages}{2421–2436}.
\newblock
\showISBNx{9781450367479}
\urldef\tempurl%
\url{https://doi.org/10.1145/3319535.3363193}
\showDOI{\tempurl}


\bibitem[\protect\citeauthoryear{Xiao, Wang, Cao, Wang, and Gao}{Xiao
  et~al\mbox{.}}{2019}]%
        {xiao2019icon}
\bibfield{author}{\bibinfo{person}{Xusheng Xiao}, \bibinfo{person}{Xiaoyin
  Wang}, \bibinfo{person}{Zhihao Cao}, \bibinfo{person}{Hanlin Wang}, {and}
  \bibinfo{person}{Peng Gao}.} \bibinfo{year}{2019}\natexlab{}.
\newblock \showarticletitle{Iconintent: automatic identification of sensitive
  ui widgets based on icon classification for android apps}. In
  \bibinfo{booktitle}{\emph{2019 IEEE/ACM 41st International Conference on
  Software Engineering (ICSE)}}. IEEE, \bibinfo{pages}{257--268}.
\newblock


\bibitem[\protect\citeauthoryear{Yeh, Chang, and Miller}{Yeh
  et~al\mbox{.}}{2009}]%
        {yeh2009sikuli}
\bibfield{author}{\bibinfo{person}{Tom Yeh}, \bibinfo{person}{Tsung-Hsiang
  Chang}, {and} \bibinfo{person}{Robert~C. Miller}.}
  \bibinfo{year}{2009}\natexlab{}.
\newblock \showarticletitle{Sikuli: Using GUI Screenshots for Search and
  Automation}. In \bibinfo{booktitle}{\emph{Proceedings of the 22nd Annual ACM
  Symposium on User Interface Software and Technology}} (Victoria, BC, Canada)
  \emph{(\bibinfo{series}{UIST ’09})}. \bibinfo{publisher}{Association for
  Computing Machinery}, \bibinfo{address}{New York, NY, USA},
  \bibinfo{pages}{183–192}.
\newblock
\showISBNx{9781605587455}
\urldef\tempurl%
\url{https://doi.org/10.1145/1622176.1622213}
\showDOI{\tempurl}


\bibitem[\protect\citeauthoryear{Zettlemoyer and St.~Amant}{Zettlemoyer and
  St.~Amant}{1999}]%
        {zettlemoyer1999visual}
\bibfield{author}{\bibinfo{person}{Luke~S. Zettlemoyer} {and}
  \bibinfo{person}{Robert St.~Amant}.} \bibinfo{year}{1999}\natexlab{}.
\newblock \showarticletitle{A {Visual} {Medium} for {Programmatic} {Control} of
  {Interactive} {Applications}}. In \bibinfo{booktitle}{\emph{Proceedings of
  the {SIGCHI} {Conference} on {Human} {Factors} in {Computing} {Systems}}}
  \emph{(\bibinfo{series}{{CHI} '99})}. \bibinfo{publisher}{ACM},
  \bibinfo{address}{New York, New York, USA}, \bibinfo{pages}{199--206}.
\newblock


\bibitem[\protect\citeauthoryear{Zhang, Ross, Caspi, Fogarty, and
  Wobbrock}{Zhang et~al\mbox{.}}{2017}]%
        {zhang2017interaction}
\bibfield{author}{\bibinfo{person}{Xiaoyi Zhang}, \bibinfo{person}{Anne~Spencer
  Ross}, \bibinfo{person}{Anat Caspi}, \bibinfo{person}{James Fogarty}, {and}
  \bibinfo{person}{Jacob~O Wobbrock}.} \bibinfo{year}{2017}\natexlab{}.
\newblock \showarticletitle{Interaction proxies for runtime repair and
  enhancement of mobile application accessibility}. In
  \bibinfo{booktitle}{\emph{Proceedings of the 2017 CHI conference on human
  factors in computing systems}}. \bibinfo{pages}{6024--6037}.
\newblock


\bibitem[\protect\citeauthoryear{Zhang, Ross, and Fogarty}{Zhang
  et~al\mbox{.}}{2018}]%
        {zhang2018robust}
\bibfield{author}{\bibinfo{person}{Xiaoyi Zhang}, \bibinfo{person}{Anne~Spencer
  Ross}, {and} \bibinfo{person}{James Fogarty}.}
  \bibinfo{year}{2018}\natexlab{}.
\newblock \showarticletitle{Robust Annotation of Mobile Application Interfaces
  in Methods for Accessibility Repair and Enhancement}. In
  \bibinfo{booktitle}{\emph{Proceedings of the 31st Annual ACM Symposium on
  User Interface Software and Technology}}. \bibinfo{pages}{609--621}.
\newblock


\end{thebibliography}

\end{document}